%% file: main.tex
\documentclass[sigconf]{acmart}

\usepackage{flushend}
\usepackage{stfloats}
\usepackage{float}
\usepackage{subfigure}
\AtBeginDocument{%
  \providecommand\BibTeX{{%
    \normalfont B\kern-0.5em{\scshape i\kern-0.25em b}\kern-0.8em\TeX}}}
\usepackage{tabularx}
\usepackage{bbding}
\usepackage{graphicx}
\usepackage{geometry}
\usepackage{amsmath}
\usepackage{makecell}
\usepackage{float}
\usepackage{color}
\usepackage[normalem]{ulem}
\usepackage{tabu}
\usepackage{hyperref}
\usepackage{xcolor}
\usepackage{appendix}

 \newcommand{\added}[1]{{\color{black} #1}}
\author{Yushang Yang}
\orcid{0009-0007-2838-608X}
\affiliation{
\institution{Studio for Narrative Spaces\\City University of Hong Kong}
\city{Hong Kong, SAR}
\country{China}}
\email{yushang.yang@outlook.com}

\author{Fanxu Meng}
\orcid{0009-0001-3551-4891}
\affiliation{
\institution{The Chinese University of Hong Kong, Shenzhen}
\city{Shenzhen, Guangdong}
\country{China}}
\email{fanxumeng@link.cuhk.edu.cn}

\author{Fiona Fui-Hoon Nah}
\orcid{0000-0002-5505-7843}
\affiliation{
\institution{School of Computing and Information
Systems\\Singapore Management University}
\city{Singapore}
\country{Singapore}}
\email{fionanah@smu.edu.sg}

\author{RAY LC}
\authornote{Correspondences can be addressed to ray.lc@cityu.edu.hk.}
\orcid{0000-0001-7310-8790}
\affiliation{
\institution{Studio for Narrative Spaces\\City University of Hong Kong}
\city{Hong Kong, SAR}
\country{China}}
\email{ray.lc@cityu.edu.hk}

\begin{document}
%%\citestyle{authoryear}

%\title["From remembering to shaping"]{"From remembering to shaping": Co-Constructing Shared Memories of Cultural Heritage in a Collaborative VR Space}
\title[From Remembering to Shaping]{From Remembering to Shaping: Narrating Shared Experiences by Co-Designing Cultural Heritage Artifacts in Collaborative VR}

\begin{teaserfigure}
  \includegraphics[width=\linewidth]{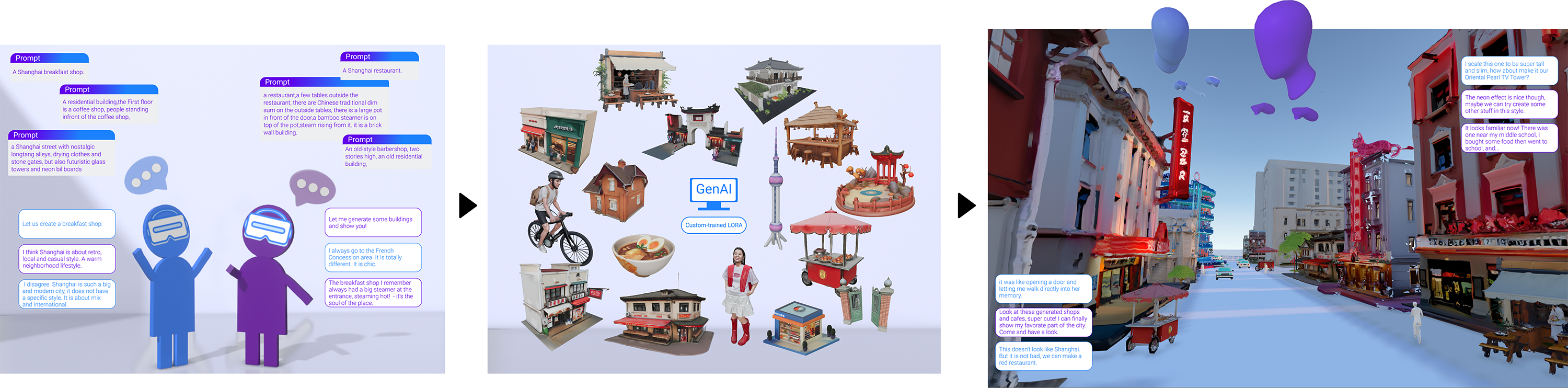}
  \caption{Participants negotiated and discussed prompts related to cultural heritage within a VR environment. They used generative AI (GenAI) to produce 3D models based on these prompts, making the visualizations concrete. Participants then manipulated the 3D models and collaboratively co-constructed memories of cultural heritage in VR space.}
  \Description{Participants negotiated and discussed prompts related to cultural heritage within a VR environment. They used generative AI (GenAI) to produce 3D models based on these prompts, making the visualizations concrete. Participants then manipulated the 3D models and collaboratively co-constructed memories of cultural heritage in VR space.}
  \label{fig:teaser}
\end{teaserfigure}

\begin{abstract}

%2026-03-31 rlc
The ways people remember and recall places reveal an invisible aspect of cultural heritage (CH), reflecting how individuals and communities relate to these places. Heritage is communal, emerging through collaboratively constructed narratives rather than individual records. To probe how people may share collective memories, we designed an immersive two-person workflow for collaboratively co-designing 3D artifacts and environments in virtual heritage locations, using Generative AI (GenAI) to instantiate these intangible memories. Observations of the co-creation process revealed that participants merged prompts and model placements when negotiating different perspectives. They used spatial operations to compose scenes, and also to express personal and embodied experiences of CH. When GenAI failed to meet their needs, participants engaged in creative appropriation, re-purposing unsatisfactory generated objects as sources of design inspiration to further shared narratives. While GenAI may have a homogenizing effect on CH expression, this work shows how people may overcome limitations in immersive collaborative workflows.

\end{abstract}

%%
%% The code below is copied from, generated by the tool at http://dl.acm.org/ccs.cfm.


%%
%% Keywords.


\maketitle
%删除copyright部分
\settopmatter{printacmref=false}
\setcopyright{none}
\renewcommand\footnotetextcopyrightpermission[1]{}
\pagestyle{plain}

%\section{Abstract}\label{sec:Abstract}
%\input{sections/00-Abstract.tex}

\section{Introduction}\label{sec:Intro}
\input{sections/DIS26_01_Intro}

\section{Related Work}\label{sec:Related Work}
\input{sections/DIS26_02_RW}

\section{Methods}\label{sec:Methods}
\input{sections/DIS26_03_Methods}

\section{Result}\label{sec:Results}
\input{sections/DIS26_04_Results}

\section{Discussion}\label{sec:Discussion}
\input{sections/DIS26_05_Discussion}

\section{Limitations}\label{sec:Limitation}
\input{sections/DIS26_06_Limitation}

\section{Design Implication}\label{sec:Design Implication}
\input{sections/DIS26_07_DI}

\section{Conclusion}\label{sec:Conclusion}

\input{sections/DIS26_08_Conclusion}

\bibliographystyle{ACM-Reference-Format} 
\bibliography{references} % references.bib 文件名

\clearpage

\appendix
\section{Appendix}
\label{sec:Appendix}
\input{sections/Appendix}

\end{document}

%% file: sections/DIS26_01_Intro.tex
% outline to the RQs 4.0:
% - purpose p1: (story belongs to more than one person) the way people remember and imagine a location is part of the invisible part of cultural heritage. but location heritage is communal, it's not just one person's record, but their collaboratively formed narratives inhabiting the place for years. can we enable people to design 3D environments to narrate these places and express their own visions of CH (imagine and remember)? collaboratively and concretely? (+refs)
% - research gap p2: current methods rely on immersive forms of visualization (AR VR) and exploration, but fail to capture people's subjective impressions, memories, desires in the form of 3D instantiations of these impressions. (+refs Miller) 
% - what we do p3: created immersive two-person platform for people to collaborately design 3D scenes using their own CH narratives, using GenAI to these environments concrete, allowing both text and image manipulation.
% RQ1: How do people design 3D scenes together in GenAI-supported immersive spaces to express their collective cultural heritage experiences?
% RQ2: How do the human-human-GenAI collaborations in designing 3D CH scenes negotiate perspective differences and resolve challenges?
% RQ3: How do spatial manipulations, as opposed to verbal descriptions, of the 3D scene affect the collaborative design process for cultural heritage expression?
% - contribution p5: colloborative design as a form of expression of invisible heritage of the places in our lives.

Cultural heritage (CH) is more than monuments or archives. It is also a cultural and social narrative process \cite{Smith2014intangibleheritage} through which people remember everyday places, imagine the future, and negotiate shared meanings that are largely invisible but vital to CH \cite{presentisinthefuture}. 
This invisible heritage is fundamentally communal \cite{He2025IRecallthePast, Fu2024BeingEroded, giaccardi2012heritage}. It is made up of personal memories, lived experiences, and future aspirations that are fragmented between individuals but become collective through storytelling and negotiation \cite{presentisinthefuture}. Furthermore, who narrates these stories, the media used for narration, and the ways in which they are narrated inevitably shape CH narratives. These narratives capture the affective and relational ties that connect people to places but are rarely reflected in official documentation \cite{jones2017wrestling, presentisinthefuture}. Despite their significance, conventional heritage approaches often privilege an Authorized Heritage Discourse (AHD)\cite{tsenova2020unauthorised}, emphasizing official histories while marginalizing unofficial memories, emotions, and local stories that remain vital to cultural identity \cite{han2014heritage, lu2023photovoice, tsenova2023locistories}. The challenge, therefore, lies not only in preservation but also in creating participatory mechanisms for communities to collaboratively author and share collective heritage narratives.

%why GenAI
\added{GenAI offers a distinct advantage for this context. It lowers the barrier to visual creation by allowing people to generate content from natural language descriptions, without the need for professional artistic or technical skills\cite{hong2023generativeaiproductdesign,han2024teams,yang_ai_2022}. This is particularly relevant for CH, where the source material is often personal memory expressed through verbal storytelling rather than technical specifications. GenAI also supports rapid iteration: participants can describe a memory, see a visual result, and refine their description in response. This iterative cycle aligns with how people recall and reconstruct memories, which are often partial and refined through conversation \cite{Brady2017memoryconstructive,Bartlett1932remembering}. In a collaborative setting, GenAI-generated artifacts can further serve as shared reference points that anchor discussion. Research in CSCW has shown that when collaborators can see and manipulate a common visual object, they establish common ground more efficiently and reduce the ambiguity inherent in purely verbal exchange \cite{Clark1991grounding, Gergle2013visualinformation}. For heritage contexts, where two people may hold different memories of the same place, having a concrete visual object to point to, critique, and revise together can turn abstract disagreements into productive negotiation.}

%The need for spatial embodiment
Yet, facilitating this co-design requires more than generating static images; it demands a spatial medium that provides the context for expressing visual designs. In the context of cultural heritage, stories are often linked to physical environments. Embedding narratives within the locations where they occur is essential for capturing spatialized sources of information. However, current GenAI applications in heritage operate primarily in 2D image formats \cite{Fu2024BeingEroded,He2025IRecallthePast}, which limit their ability to represent place-based meaning, as physical environments are defined by spatial relationships and scale, not merely visual textures. Without translating abstract memories into concrete 3D forms, participants may be constrained in their ability to express the spatial relationships that shape the story of a place. 

%Active Engagement and Collaboration in VR
While immersive technologies provide 3D immersion, most CH applications remain expert-led \cite{bozzelli2019arkaevision,innocente2023immersiveXRframework} or focused on single-user exploration \cite{innocente2023immersiveXRframework,ShinandWoo2023HowSpaceIsTold}, positioning people as viewers of pre-defined stories rather than active co-creators of heritage meaning. This gap highlights an opportunity to support collaborative co-creation practices that enable people to reframe and collectively gather stories that reflect their subjective impressions and recollections. 

Importantly, the spatial affordances of VR, including the ability for multiple users to collaboratively interact with 3D objects, remain underexplored in heritage negotiation. Notably, spatial action serves as a form of non-verbal negotiation that complements verbal discourse \cite{Chang2017EvaluatingtheEffectofTangibleVR}. While recent work has explored GenAI-supported collaboration in immersive settings \cite{miller2025eliciting}, it primarily focuses on visual editing to elicit memories, rather than using spatial manipulation to resolve conflicts or align narratives. There is a need to explore how integrating spatial actions with GenAI can facilitate the negotiation and co-creation of heritage spaces. To enable this collaborative storytelling in CH, we propose a workflow that integrates the generative power of AI with the spatial affordances of VR. By instantiating abstract stories as tangible 3D objects, the workflow positions virtual space as a site of negotiation in which participants visually construct, negotiate, and align shared narratives.

To investigate how the spatial affordances of VR and GenAI can facilitate collective heritage narratives, we structure our inquiry around three research questions:

RQ1: How do people design 3D scenes together in GenAI-supported immersive spaces to express their collective cultural heritage experiences?

RQ2: How do the human-human-GenAI collaborations in designing 3D CH scenes negotiate perspective differences and resolve challenges?

RQ3: How do spatial manipulations, as opposed to verbal descriptions, of the 3D scene affect the collaborative design process for cultural heritage expression?

%What we do:
To address the research questions, we designed a participatory approach that empowers people to author their own narratives using an immersive two-person workflow for collaboratively co-constructing cultural memories. \added{We ground this study in Shanghai's Haipai culture, a context shaped by over a century of East-West cultural exchange, which provides a site where participants naturally hold divergent perspectives on the same urban spaces, creating conditions for observing how collaborative negotiation unfolds.} In this system, participants entered a shared VR environment, where they created and built their visions of cultural spaces. They then discussed their ideas and provided prompts to GenAI to create 3D models about cultural locations; these models could then be directly manipulated in the shared immersive space (e.g., moving and scaling). 

The study identified patterns in collaborative design practice that extend beyond the placement of objects. Participants used embodied spatial operations not only to arrange scenes but also as a mechanism for implicit negotiation and embodied recall of memories. For instance, repositioning objects or scaling could serve as a non-verbal strategy to resolve conflicting cultural perspectives without confrontation, while spatial distance was used to express the ineffable atmosphere of memory. Furthermore, when GenAI produced unexpected outputs, participants engaged in creative appropriation, re-purposing unsatisfactory objects for their shared narrative (e.g., transforming an unintended building into a sculpture). These observations highlight a dynamic interplay in which spatial interaction functions as a distinct design vocabulary alongside verbal discourse.

This study framed collaborative design with GenAI in a VR space as an active form of expressing the invisible heritage of the places in our lives, showing how cultural heritage can be represented as a dynamic, immersive co-design process. Together with GenAI, immersive space is not only a visualization tool but also functions as a platform for collaborative storytelling. We demonstrate how the co-design process itself allowed people to visualize memories, negotiate perspectives, and envision future aspirations. We present the design workflow and provide insights into human-human-GenAI collaboration by showing how spatial interaction and generative tools can be combined to support communal ways of authoring and shaping cultural heritage. Ultimately, by designing imagined artifacts in the virtual world, people can effectively articulate the CH of the real world.

%% file: sections/DIS26_02_RW.tex
\subsection{Representing Cultural Heritage: From Physical Artifacts to Human Stories}
The representation of CH has undergone a significant shift in recent decades, moving from a primary focus on the preservation of physical artifacts to an increasing emphasis on human-centered storytelling and collective memory \cite{gentrification2019storytelling}. Early work in CH research was primarily concerned with material preservation and reproduction of tangible entities, such as historical buildings, archaeological remains, and museum objects \cite{avrami2009heritage,jokilehto2017history}. While such approaches contributed to safeguarding material culture, they often reduced heritage to static objects of observation, neglecting the emotional, social, and narrative dimensions that connect communities to their cultural past. As a result, audiences frequently remained passive observers rather than active participants in meaning-making \cite{Smith2006usesofheritage}. 

The development of digital technologies introduced new opportunities for CH representation. Virtual and augmented reality have been widely adopted for applications such as virtual museums~\cite{cao_dreamvr_2023}, immersive heritage reconstructions, interactive exhibitions~\cite{zhou_retrochat_2025}, and history heritage pervasive games \cite{bekele2018survey,carrozzino2010beyond,elin2014mystery}. However, these implementations often remain limited to reproducing physical sites or generating static 3D reconstructions \cite{historyis3d}, which, although visually compelling, rarely leverage the full potential of immersive media to enable interaction, participation, and collaborative meaning construction \cite{champion2015critical}. 
%shorten why 3d, add Gap, and what we do
The key question of “why 3D?” thus becomes central: beyond visual immersion, 3D spaces enable embodied interaction \cite{Paul2001WheretheActionis} and multi-perspective interpretation within shared environments. \added{Yet few heritage systems have used these spatial affordances for collaborative meaning-making rather than passive viewing.}

Recent work has increasingly emphasized the humanistic dimensions of heritage, recognizing that the essence of heritage lies not only in objects but also in lived experiences, local perceptions, and collective memory \cite{ciolfi2002Designing,giaccardi2012heritage}. A critical gap in heritage digitization concerns the lack of “human data”, which consists of the voices, stories, and affective connections that underpin cultural significance \cite{kalay2008new}. Current approaches often rely on interviews or curated testimonials, which flatten complex, situated experiences into static records. Such approaches highlight a methodological limitation: existing representations tend to “display” human data rather than “enact” them through participatory and spatialized practices. Spatialization has been proposed as a way to make such data negotiable among collaborators \cite{Paraschivoiu2025CraftingCities}. \added{However, this idea has seen limited application in heritage contexts where communities hold competing memories of the same place.}

Research indicates that spatial layout and physical environments influence memory formation and transmission \cite{Krumpen2021CollaborativeVR, Ciolfi2013DesigningForEngagement}. Open spatial arrangements can promote participant interaction, whereas specific configurations guide the sharing and interpretation of cultural narratives. Sensory qualities of cultural heritage and environments, including tactile and visual cues, further enhance memory recall and narrative engagement \cite{Paul2001WheretheActionis, Malegiannaki2020NarrativeGameHeritage}. \added{However, less is known about how people themselves arrange and manipulate spatial elements to construct and negotiate their own collective memories.}

In summary, while the paradigm in digital heritage has shifted from "objects to people," research often still positions users as the audience of stories rather than active co-creators. The critical gap, therefore, is not the lack of humanistic focus but the lack of mechanisms for collaborative authorship and spatialized co-creation. Addressing this gap requires moving beyond preservation or exhibition toward an understanding of digital heritage as a living process of negotiation. \added{Our research contributes to this direction by empowering community members to author their own heritage narratives through spatial co-design, rather than consuming stories curated by experts.}

\subsection{GenAI as a Narrative Medium}
GenAI has been widely applied in the restoration of cultural heritage, as well as visualization and creative expression, such as text generation \cite{assael2022restoring} and image generation \cite{Yuan2025huayao, liu_dreamscaping_2024,liu_salt_2025}. GenAI addresses the skills gap by lowering entry barriers, enabling novices to create visual content through natural language interaction. This capability empowers users to draw inspiration and articulate their narratives without requiring professional artistic skills \cite{hong2023generativeaiproductdesign,lc_together_2023}. Furthermore, GenAI’s ability to generate novel and unexpected outputs can inspire novices, while its training on diverse data sources allows it to shape personal narratives and offer distinctive visual perspectives that prompt deeper reflection \cite{chiou2023designingwithai}. In CH dissemination, visual outputs can convey complex ideas more effectively than text alone, fostering introspection and engagement \cite{liu2020evaluatingdigitalinterpretation,watson2010VisualityandPast}. The cultural background and perspective of each participant will shape the visual narrative and collective memory \cite{Benjamin1968WorkofArt}. However, the application of GenAI in cultural narrative remains constrained by two major limitations in current research paradigms.

\textit{Lack of attention for the critical role of 3D space for collaborative meaning-making.} The existing GenAI-assisted cultural heritage narrative tools mainly focus on 2D images, such as text-to-image generation by Midjourney. In one study, users selected a CH they had an emotional connection with, imagined possible threat scenarios for the heritage site in the future, and used Midjourney text prompts to generate scene images \cite{Fu2024BeingEroded}. Although effective for individual expression, these 2D artifacts are primarily designed to be viewed as static outputs. However, CH is inherently spatial; a place is not just a view to be observed, but a space to be inhabited with its own subculture~\cite{fu_i_2023}. Compared to 3D environments, 2D artifacts lack support for real-time collaborative manipulation—a capability critical for negotiating shared understanding among users \cite{Paraschivoiu2025CraftingCities}. The recent emergence of 3D GenAI models, such as Tencent Hunyuan \cite{zhao2025hunyuan3d}, Tripo AI \cite{Zhao2025RevivingMuralArt}, and NeRF \cite{mildenhall2021nerf}, offers a path beyond this limitation, enabling the generation of interactive 3D models that can be placed and manipulated in shared digital spaces for narrative interventions \cite{gong_if_2025}. 

\textit{Research has largely focused on single-user scenarios while ignoring multi-user collaboration.} Existing research on intangible cultural heritage assisted by GenAI mainly focuses on single-person scenarios \cite{wang2025harmonycut, adiba2024exploring}. However, cultural heritage is inherently collective \cite{silberman2012collective}. When multiple users co-create, differences in interpretation, intention, and cultural background inevitably arise, which leads to the need for negotiation and reconciliation among participants. These multi-user contexts introduce new complexities for both users and GenAI systems: how to negotiate the differences among multi-users, handle the potential errors of GenAI, and define the role of GenAI. How to share AI-generated 3D artifacts as collaborative objects among multiple parties in immersive spaces remains an unsolved problem.

Furthermore, even in studies that combine VR and GenAI, the integration is often functional and modular rather than deeply synergistic. For instance, a study on the restoration of Dunhuang murals integrated the VR and GenAI technologies to form an efficient technical pipeline, from segmenting and classifying murals to fixing the picture with GenAI, and finally assembling the generated objects in Unity 3D to create an immersive VR scene \cite{Zhao2025RevivingMuralArt}. Another study that combines VR and GenAI was dedicated to transforming the cultural heritage of performing arts into spatial interactive experiences \cite{wang2025temporal}. However, in these approaches, GenAI and VR each play their own roles; GenAI was used to separate the audio of traditional operas and generate visual environments, while VR provided the display and core interactive fields.

Overall, these approaches are often either technology-driven and centered on the goal of efficient restoration or treat GenAI and VR as relatively separate functional modules (GenAI for generation, VR for display). A significant research gap remains in exploring how to deeply integrate both technologies into a unified socio-technical system that supports real-time collaboration, negotiation, and co-creation among multiple users. In particular, the question of how GenAI can function as a collaborative role rather than merely a production tool to mediate interpersonal interactions in immersive spaces has not been thoroughly investigated. \added{We address this gap by designing an immersive workflow 
in which GenAI serves not only as a production tool but as a mediator in the collaborative construction of heritage narratives within shared 3D space.}

\subsection{Spatial Interaction and Collective Memory}

The gaps we have identified in heritage representation and GenAI tools point to a unified challenge: moving beyond static, single-user consumption towards dynamic, collaborative co-creation. We argue that the solution lies in understanding spatial interaction and supporting the construction of collective memory.

Foundational theories in CSCW posit that collective memory is actively constructed through shared spatial practices. A key idea is the difference between "space" (just a physical area) and "place" (a space that has meaning because of what people do there) \cite{Harrison1996Re-place-ingSpace}. This perspective suggests that people create ‘place’ and shared understanding through their actions: by moving objects, walking through areas, and looking at things from different views together. However, many heritage applications, although visually compelling, have been criticized for failing to move beyond static display, lacking the tools for genuine interaction and collaborative meaning construction \cite{Ciolfi2013TheCollaborativeWorkofHeritage}.

Immersive technologies like VR offer a powerful environment for such collaborative place-making. Unlike traditional digital interfaces, VR enables embodiment and shared presence, allowing participants to interact with digital artifacts and each other in a manner that closely approximates real-world collaboration \cite{Paul2001WheretheActionis}. The ability to walk around a model, point to a feature, or stand side-by-side to discuss its placement are not trivial features; they are the basic ways that people collaborate and create meaning together. Empirical studies in social VR \cite{Freeman2022WorkingTogetherApartthroughEmbodiment} provide direct evidence, showing how embodied interaction facilitates more natural collaboration.

These insights define a clear research direction. For digital heritage, the challenge lies in moving beyond the presentation of stories to enabling shared experiential engagement \cite{Ciolfi2013TheCollaborativeWorkofHeritage}. For GenAI, the challenge is to move beyond solitary, 2D content creation to collaborative, 3D spatialized workflows. Our study addresses this confluence by designing and deploying an immersive, two-person workflow that uses GenAI for the co-construction of cultural heritage memories. \added{By integrating GenAI into a shared VR 
environment, we examine how collaborative narratives unfold when participants are empowered to not only interact within a space but also actively co-create that space itself, addressing the core challenge of merging GenAI’s narrative capabilities with VR’s spatialized affordances to support collective memory construction.}

%% file: sections/DIS26_03_Methods.tex
Our study investigates how immersive VR environments, combined with generative AI, can support collaborative reconstruction of cultural heritage spaces from multiple perspectives. In pairs, participants co-created a 3D street scene representing their impressions of Shanghai’s central district, placing AI-generated models directly into a shared VR environment.

\begin{figure} [tp]
 \centering
 \includegraphics[width=1\linewidth]{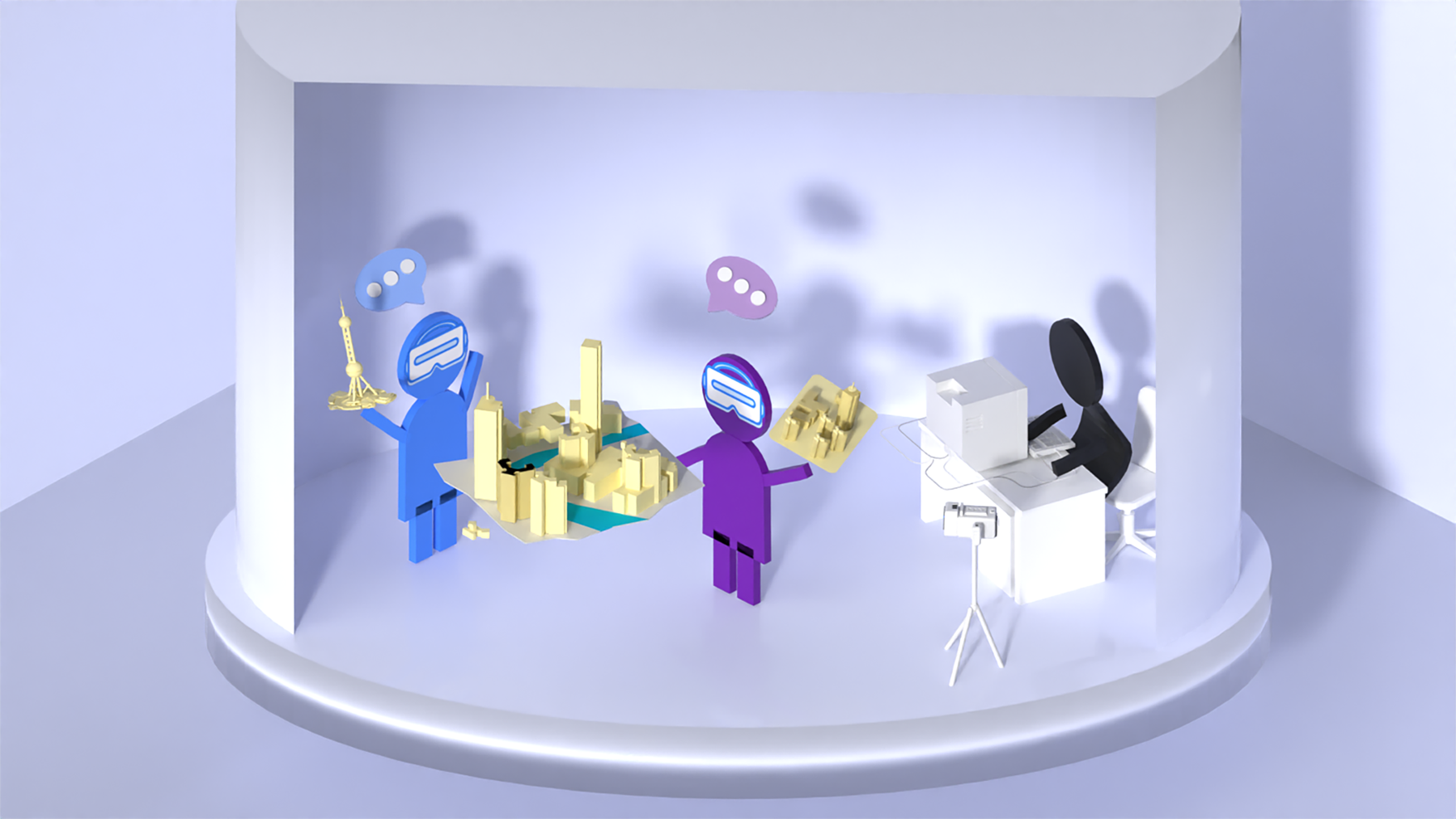}
 \caption{An illustration of the study setup. Two participants, each wearing a VR headset, collaboratively interacted with AI-generated 3D models in a shared virtual space. A researcher, external to the VR environment, facilitated the process and recorded data.}
 \Description{An illustration of the study setup. Two participants, each wearing a VR headset, collaboratively interacted with AI-generated 3D models in a shared virtual space. A researcher, external to the VR environment, facilitated the process and recorded data.}
 \label{fig:3-setup}
\end{figure}

\subsection{Participants} 

We recruited 18 participants (9 pairs) through social media outreach and personal referrals, encompassing diverse professional backgrounds such as software development, design, law, and marketing (see Fig. 3). Participants were randomly assigned into pairs. Demographic and background information were collected through a pre-study questionnaire. To ensure participants had a personal connection to the study context, we recruited individuals who had lived, worked, or studied in Shanghai for at least 3 years in the past 10 years, ensuring they possessed a foundational familiarity with the city. Participants volunteered for the study.

\begin{figure*}[!t]
 \centering
 \includegraphics[width=0.6\linewidth]{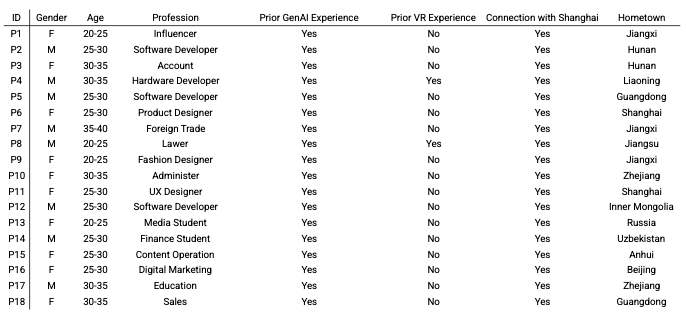}
 \caption{Demographic table}\label{Demographic}
 \Description{Demographic table}
\end{figure*}

\subsection{Study Context}

To effectively observe collaborative design processes, we required a cultural context that naturally elicits diverse and potentially conflicting individual perspectives. \added{Few cities have sustained direct contact between Eastern and Western cultural forms as a defining feature of their built environment. Unlike static or mono-cultural heritage sites, Shanghai's Haipai culture serves this purpose well. Shaped by over a century of East-West exchange following Shanghai's opening as a treaty port in 1843, Haipai represents a sustained fusion of Chinese and Western aesthetics embedded in everyday architecture and urban life\cite{greenspan2014shanghai,Liu2023Haipai}.} This diversity is essential to inspire participants to express their own visions and negotiate differences, rather than simply settling for a common consensus. 
Furthermore, Shanghai's dramatic urban transformation \cite{Yin2011urbanchangesofshanghai} has led to rapid changes in the city's landscape and living experiences, potentially causing people to have different impressions of the same neighborhood. This complexity in cultural and urban environments allows participants to hold distinct interpretations of the site, creating opportunities to observe how the system supports complex cultural expression and negotiation. \added{ We prioritize depth of cultural understanding, a position consistent with the value of non-Western, locally situated research\cite{Linxen2021howweirdischi,Qadri2025nonwesternartworlds}.}

\subsection{Procedure}

Participants were co-located in the same physical space to foster natural conversation and allow for the observation of non-verbal cues (e.g., gestures, laughter), providing a richer dataset than remote collaboration. The VR environment they shared contained a simplified 3D base map of central Shanghai, providing only geographic references to avoid biasing creative decisions.

The session for each pair was 120 minutes, structured as follows:
\begin{itemize}

\item Training (20 mins): Participants were familiarized with the VR controls and our system through a brief demonstration with example prompts showing the whole process from prompts to VR manipulation, such as how to drag, zoom, and change the viewport.
\item Collaborative Creation (80 mins): The core task was open-ended: “Construct a street in Shanghai’s central district as you imagine it.” Participants engaged in iterative cycles of generation, placement, and discussion with no restrictions on style or strategy.
\item Semi-structured Interview (20 mins): A post-task interview covered themes of immersion, collaboration, and their experience with the AI.
\end{itemize}

\begin{figure*}[!h]
 \centering
 \includegraphics[width=0.9\textwidth]{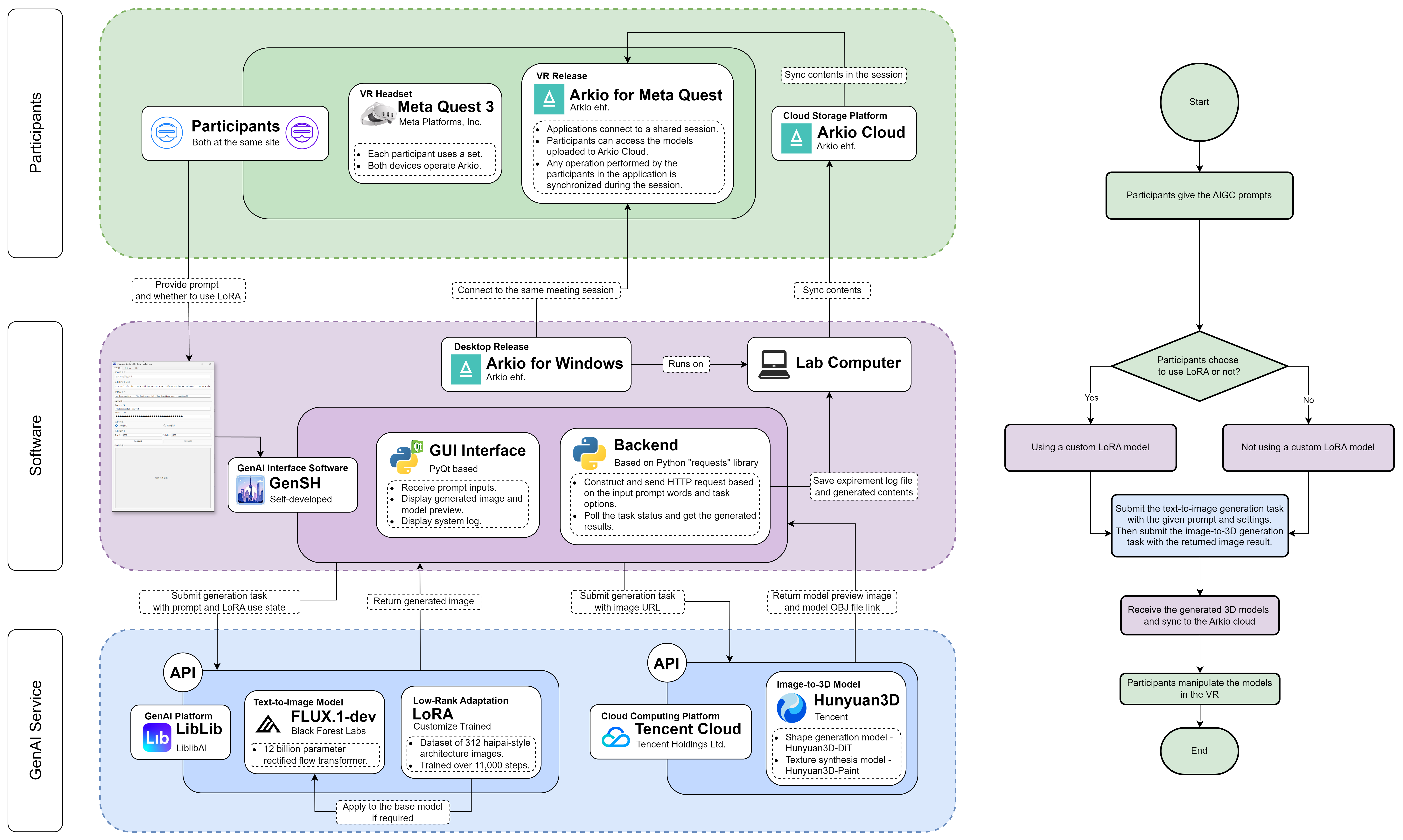}
 \caption{System design and workflow} 
 \Description{System workflow}
 \label{fig:system}
\end{figure*}

\subsection{System Design}

% == This figure was from the result section ==
% \begin{figure*}[tb]
%     \centering
%     \includegraphics[width=1\linewidth]{figures/4.0 summary.png}
%     \caption{Screenshots of participants' first-person perspective, the process of creating and editing the generated model in space, and the streets that participants create in VR with GenAI.}
%     \Description{Screenshots of participants' first-person perspective, the process of creating and editing the generated model in space, and the streets that participants create in VR with GenAI.}
%     \label{fig:summary}
% \end{figure*}
% == This figure was from the result section ==

%Intro of VR
We used Meta Quest 3 headsets to host a shared VR environment built on Arkio, a commercial spatial design platform. This platform enabled participants to collaboratively place, resize, and manipulate 3D models using standard VR controllers and provided key environmental controls like custom day-night cycling with dynamic lighting.

%Intro of AI used. Minor grammar correction
The primary workflow used text-to-image-to-3D generation, and our generative backend employed two models: FLUX.1-dev text-to-image model \cite{flux2024} developed by Black Forest Labs (BFL) and deployed on the LibLib platform, and Hunyuan3D \cite{zhao2025hunyuan3d}, an image-to-3D model by Tencent deployed on Tencent Cloud. Both models were connected via API interfaces integrated into our GenSH software to accelerate generation time. Flux was selected for its high fidelity in generating images from natural language, while Hunyuan3D was chosen for its high precision and stability in 3D modeling.

%Intro of LoRA 
To address the base models' limited understanding of local architecture, we implemented a culturally-customized design strategy by training a custom LoRA model on the Haipai (Shanghai local architecture style). Using a dataset of 312 images trained over 11,000 steps, this component served as a technical intervention to bridge the gap between GenAI knowledge and specific cultural heritage. In our study design, this model was not treated as a formal experimental condition for comparison, but rather as an optional empowerment tool. Participants were informed that they could request to use this specialized model at any point, particularly if they felt the standard model failed to capture the specific architectural style they envisioned.

%Intro of tool
A supported tool GenSH was developed to provide one-click functions to generate 2D images or 3D models from participant-provided prompts and the option of LoRA. The AI-generated 3D models could then be imported into the shared VR scene for participants to use and edit. 
%The WOZ
Current voice-to-text systems often capture extraneous speech, such as filler words, conversational asides, and thinking-aloud protocols, alongside the intended prompts. If fully ingested, this noise can degrade AI generation quality and misrepresent participant intent. Conversely, typing within a VR environment is significantly slower than physical typing, and virtual keyboards disrupt the sense of presence. To ensure interaction fluidity and mitigate the latency and error rates of current VR text input, we employed a Wizard of Oz (WOZ) protocol. Participants gave the prompts of objects/buildings verbally, and a researcher inputting the prompts verbatim into the GenSH tool, which would generate 3d models for participants to modify in the VR space. Participants could verify the prompts and request corrections anytime during the study. This design choice prioritizes the participants' sustained immersion and creative flow over technical automation.
%Optimization Features
To ensure consistent generation quality, the GenSH interface automatically appended technical modifiers (e.g., view angles like ``45-degree orthogonal", or the trigger word ``Shanghai building" for LoRA activation) to user inputs. Participants retained full control to modify these presets. This feature helped participants to focus on design rather than technical efforts.

\subsection{Data Collection and Analysis}

We collect our data from multiple perspectives:
\begin{itemize}

\item VR environment recording (screen capture of the shared scene).
\item Audio recording of all verbal exchanges.
\item System logs of AI generations, including timestamps and prompts.
\item Photographic and video documentation of the physical session.
\item Researcher observation notes taken in real time.
\end{itemize}

Semi-structured interviews after the task covered immersion, spatial expression, collaboration, AI interaction, and overall impressions. \added{We analyzed the qualitative data using Thematic Analysis \cite{Braun2006}. Two researchers independently familiarized themselves with the full dataset, comprising interview transcripts, observation notes, and session recordings, before generating initial codes. The researchers then met to compare their codes, discuss interpretations, and collaboratively develop a set of candidate themes. Where interpretations differed, disagreements were resolved through discussion until consensus was reached. The resulting themes were reviewed against the full dataset to ensure they accurately represented the range of participant experiences.}

Interview analysis focused on participants’ reflections on immersion, collaboration, and their perception of AI’s role. Observation analysis coded non-verbal coordination, spatial manipulations, and physical-virtual interactions, not always verbalized. Content analysis systematically categorized all AI-generated 3D models by style (e.g., haipai, modern, hybrid). A table is provided in the appendix.

\subsection{Ethics}

Our research protocol was approved by the university's institutional review board (IRB). Before the study began, all participants provided written informed consent. We informed each person that their participation was voluntary and that they could stop the experiment at any time. To minimize discomfort from using the VR equipment, we instructed participants to report any feelings of nausea or unease immediately. If a participant reported discomfort, they were offered a break or the option to end their participation. In addition, VR devices will be disinfected after each use. All the experiment data was anonymized during transcription and securely stored for academic research use only.

%% file: sections/DIS26_04_Results.tex
%Minor grammar adjustments only

\begin{figure*}[!t]
    \centering
    \includegraphics[width=1\linewidth]{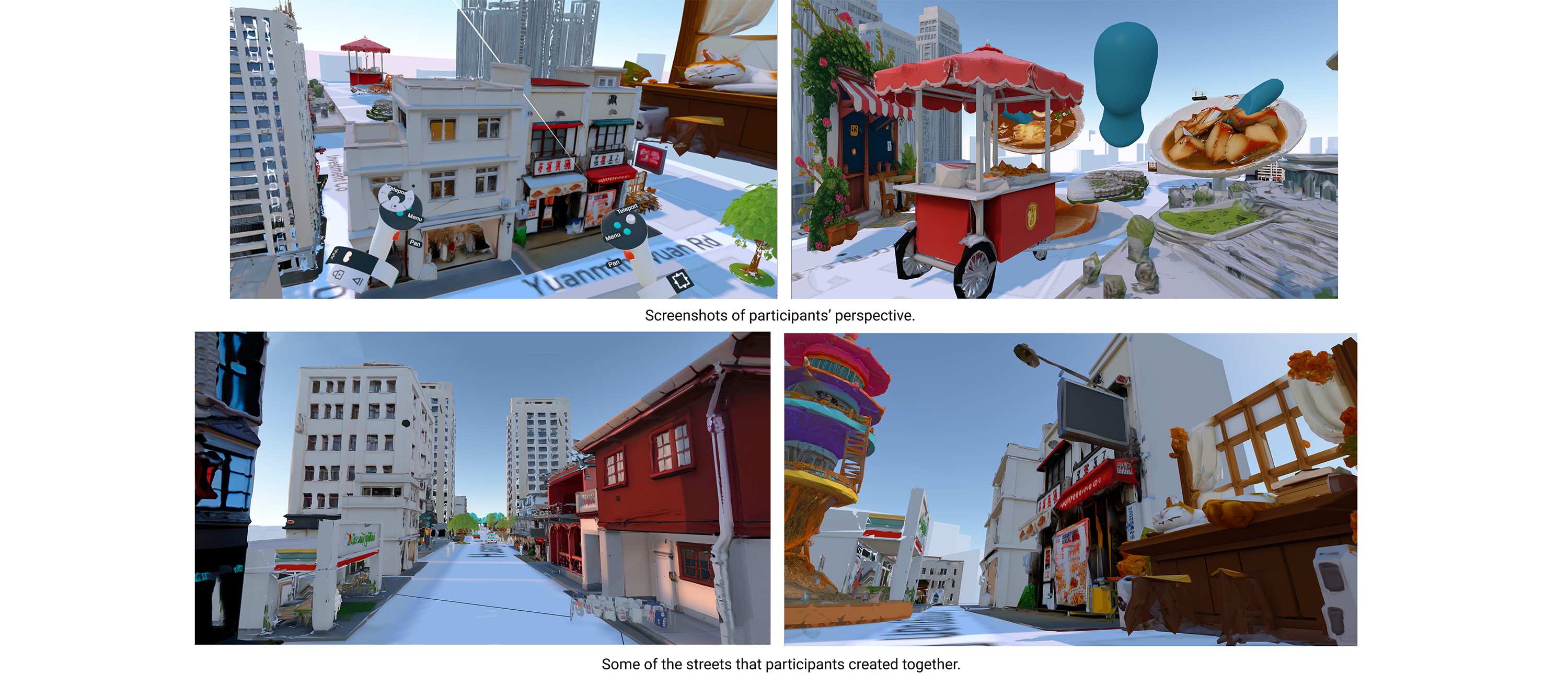}
    \caption{Screenshots of participants' first-person perspective, the process of creating and editing the generated model in space, and the streets that participants create in VR with GenAI.}
    \Description{Screenshots of participants' first-person perspective, the process of creating and editing the generated model in space, and the streets that participants create in VR with GenAI.}
    \label{fig:summary}
\end{figure*}
% == This figure was moved to just under the start of system design ==

\added{Our analysis draws on in-session observations, system logs, and post-study interviews. The findings are organized into three parts. Section 4.1 examines how participants used spatial manipulation of 3D objects-both as a design tool and as a form of non-verbal negotiation-to co-construct shared heritage scenes. Section 4.2 describes how participants navigated differences in cultural knowledge and aesthetic preferences during collaboration. Section 4.3 reports how participants responded when GenAI outputs did not meet their expectations, including strategies for repurposing and adapting unexpected results.}

\subsection{Process of Co-Constructing Shared Memories with GenAI in VR}
\subsubsection{Creating Shared Memories with AI-Generated Objects} 

\begin{figure*}[!h]
    \centering
    \includegraphics[width=0.9\linewidth]{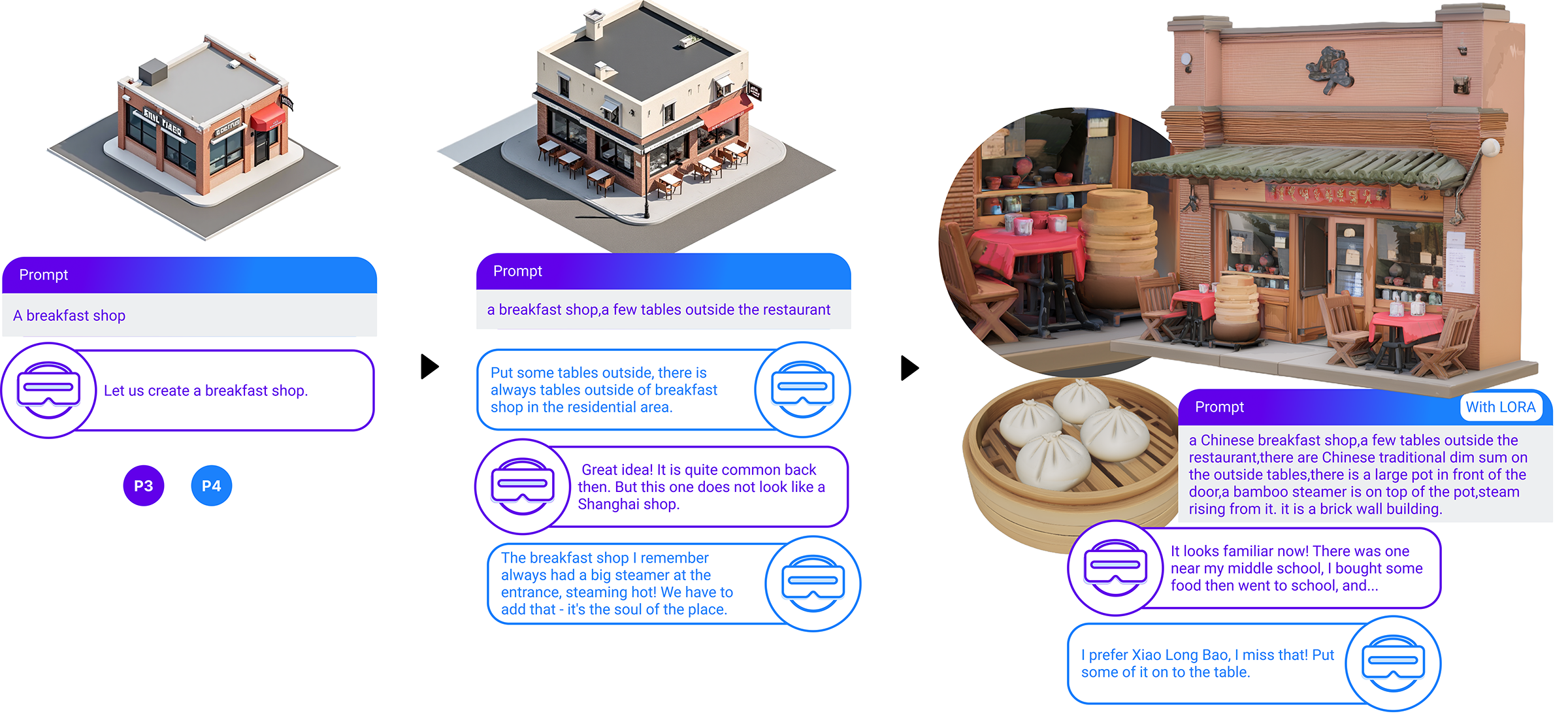}
    \caption{P3's initial generation of a breakfast shop (left) prompted P4 to recall a specific memory of a steamer at the entrance. They then collaborated to add this detail (right), creating a richer, shared scene.}
    \Description{P3's initial generation of a breakfast shop (left) prompted P4 to recall a specific memory of a steamer at the entrance. They then collaborated to add this detail (right), creating a richer, shared scene.}
    \label{fig:breakfast}
\end{figure*}

In our multi-user VR environment, we observed participants using AI-generated 3D objects to visualize their personal and collective memories. One participant's generated object could lead their partner to recall and describe additional details. For example, when P3 generated a simple breakfast shop, P4 immediately recalled a specific detail: "The breakfast shop I remember always had a big steamer at the entrance, steaming hot! We have to add that, it's the soul of the place." They then collaboratively refined the prompt to include this detail. Once the objects were generated, the pair shifted to arranging them to match the lively atmosphere of a morning market. Initially, the shop was too spacious. To address this, P4 dragged the models tightly together to create a sense of congestion. Walking through this condensed arrangement, P3 noted, "Now it feels right! It feels like the crowded lanes I remember." In the post-study interview, P4 reflected that the final scene felt "richer and more detailed than what I could have imagined alone." (figure \ref{fig:breakfast})
%fig6
%fig7
\begin{figure*}[tp]
    \centering
    \includegraphics[width=0.9\linewidth]{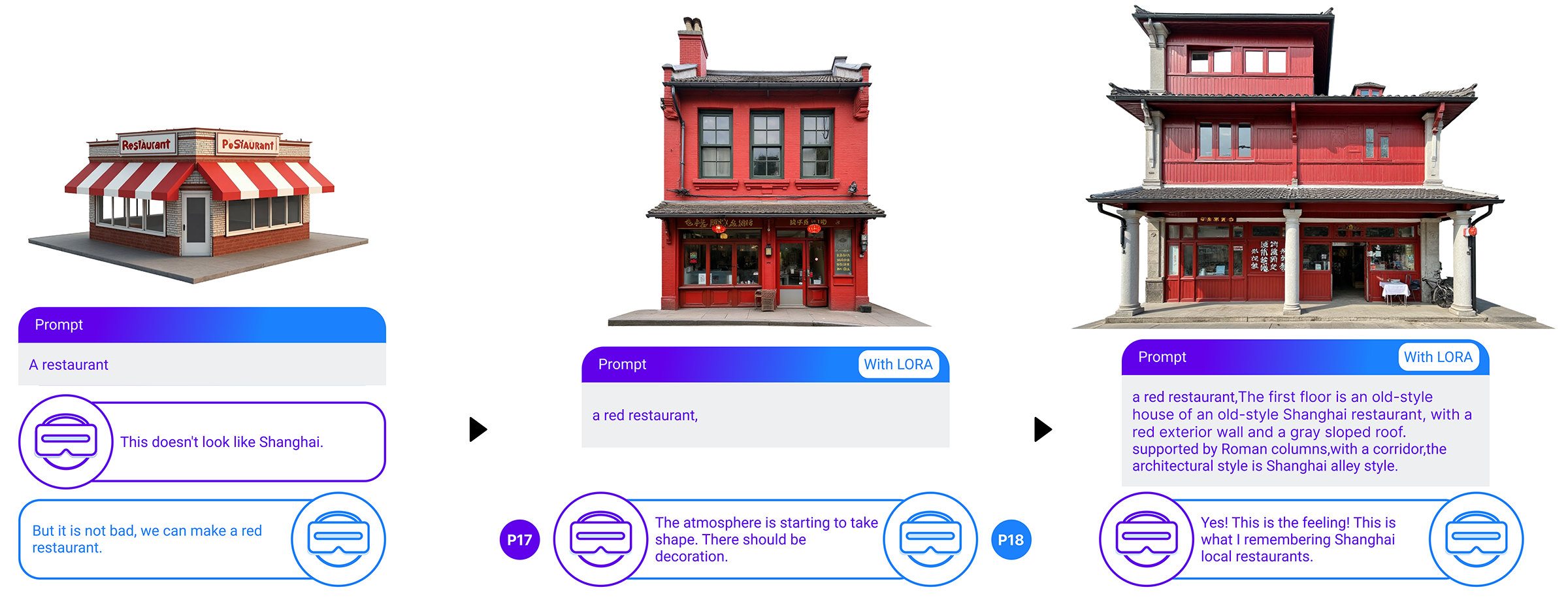}
    \caption{The fine-tuned LoRA model streamlined the creative process. Participants could use simpler prompts (e.g., "a red restaurant") to generate culturally specific architecture, which then served as a foundation for adding more detailed prompts.}
    \Description{The fine-tuned LoRA model streamlined the creative process. Participants could use simpler prompts (e.g., "a red restaurant") to generate culturally specific architecture, which then served as a foundation for adding more detailed prompts.}
    \label{fig:loraprompt}
\end{figure*}

When participants chose to use the custom LoRA model, we observed that the generated content exhibited a stronger Shanghai architectural style. This reduced their need to elaborate on every architectural detail in their prompts. For instance, P17 and P18 initially used the general Flux model and were disappointed, with P17 commenting, "This doesn't look like Shanghai." They then switched to the LoRA model. P18 provided a simple prompt: "Help me generate a red restaurant." The system generated a two-story building with a retro Shanghai style, which met their expectations. They continued to build upon this output, adding more specific details in subsequent prompts. After building out the scene, P18 stated in the post-study interview that they "felt as if we had been transported back to that street." (figure \ref{fig:loraprompt})

%fig8
\begin{figure*}[!h]
    \centering
    \includegraphics[width=0.9\linewidth]{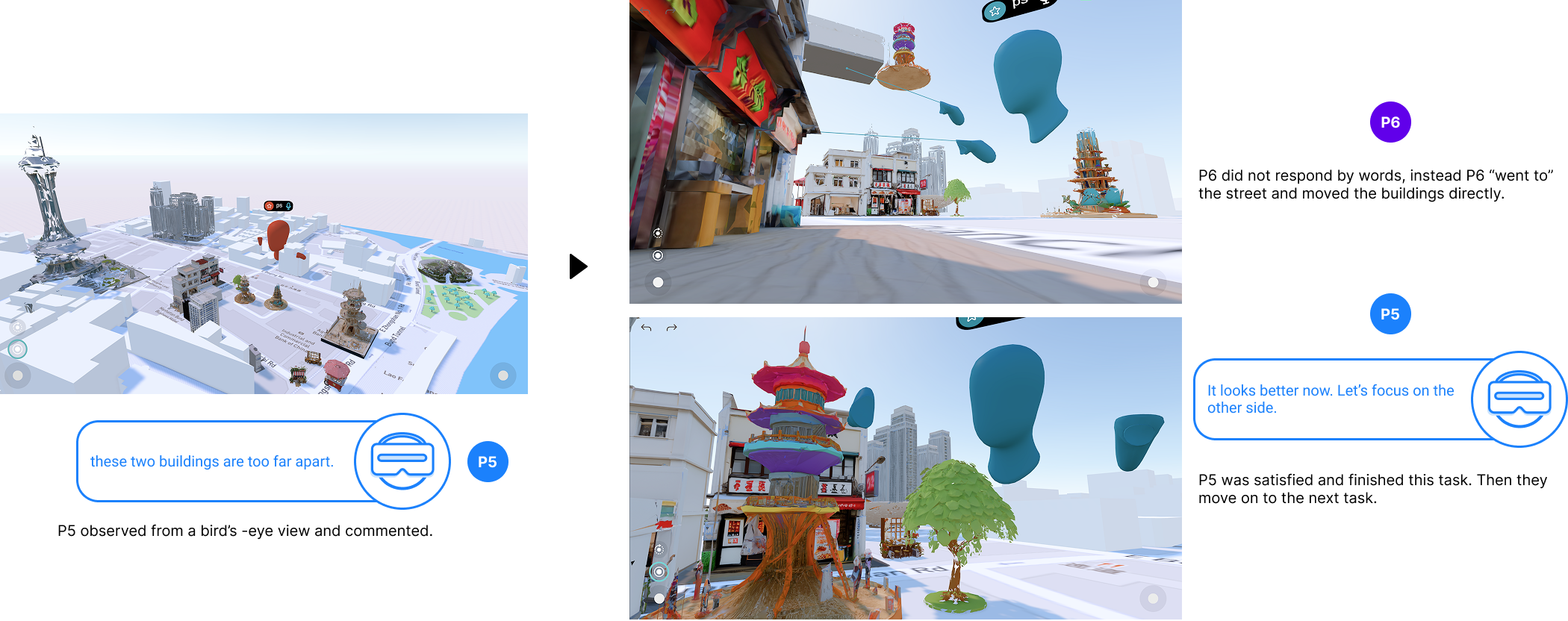}
    \caption{An example of "action as dialogue." In response to the verbal comment, a participant uses direct manipulation to physically drag the buildings closer, immediately resolving the issue and opening up a new design possibility.}
    \Description{The fine-tuned LoRA model streamlined the creative process. Participants could use simpler prompts (e.g., "a red restaurant") to generate culturally specific architecture, which then served as a foundation for adding more detailed prompts.}
    \label{fig:moveaction}
\end{figure*}
%fig9
\begin{figure*}[tp]
    \centering
    \includegraphics[width=0.9\linewidth]{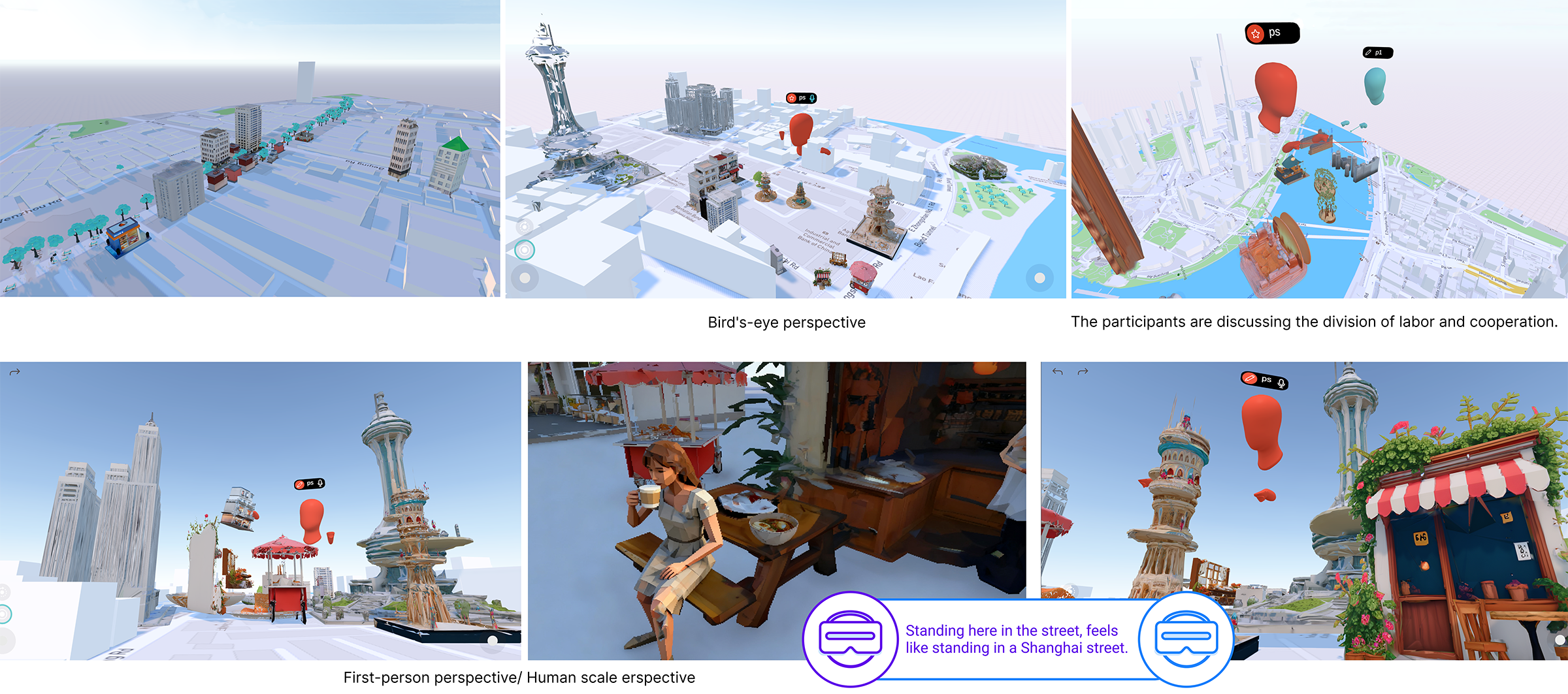}
    \caption{Participants strategically switched between perspectives.}
    \Description{Participants strategically switched between perspectives.}
    \label{fig:perspective}
\end{figure*}

\subsubsection{Negotiating by Directly Manipulating Objects in 3D Space}  
%修改
We observed that participants frequently used direct manipulation of 3D models to resolve spatial disagreements, often bypassing verbal discussion. In one instance, as in figure \ref{fig:moveaction}, P5 verbally noted that "these two buildings are too far apart." In response, P6 physically grabbed the models and moved them closer. P5 then replied, "It is better now," and the pair moved to the next task. Similarly, when P12 felt the sign created by P11 clashed with the historical gate they had built, instead of verbally rejecting the idea, P12 simply grabbed the sign and dragged it from the center to the side. P11 accepted this new arrangement without objection. This spatial adjustment allowed the conflicting element to exist without disrupting the primary narrative. In another case, P7 and P8 recreated a narrow street from memory. After switching to a first-person perspective to walk through it, P7 remarked, "This feels authentic... But it's too oppressive to walk in." In response, P8 immediately grabbed the building models and dragged them apart to widen the street. After observing the change, P7 asked, "So, can we 'repair' this memory? We have the power now to create and make it better," and proposed designing a plaza. The pair subsequently began discussing new design elements for the widened space. Later in the session, P8 reflected on this change:\begin{quote}
    \textit{"Look how interesting our goal has changed after widening the street. We wanted a narrow street, but now we have an avenue."}
\end{quote}

%fig10
\begin{figure*}[!h]
    \centering
    \includegraphics[width=0.9\linewidth]{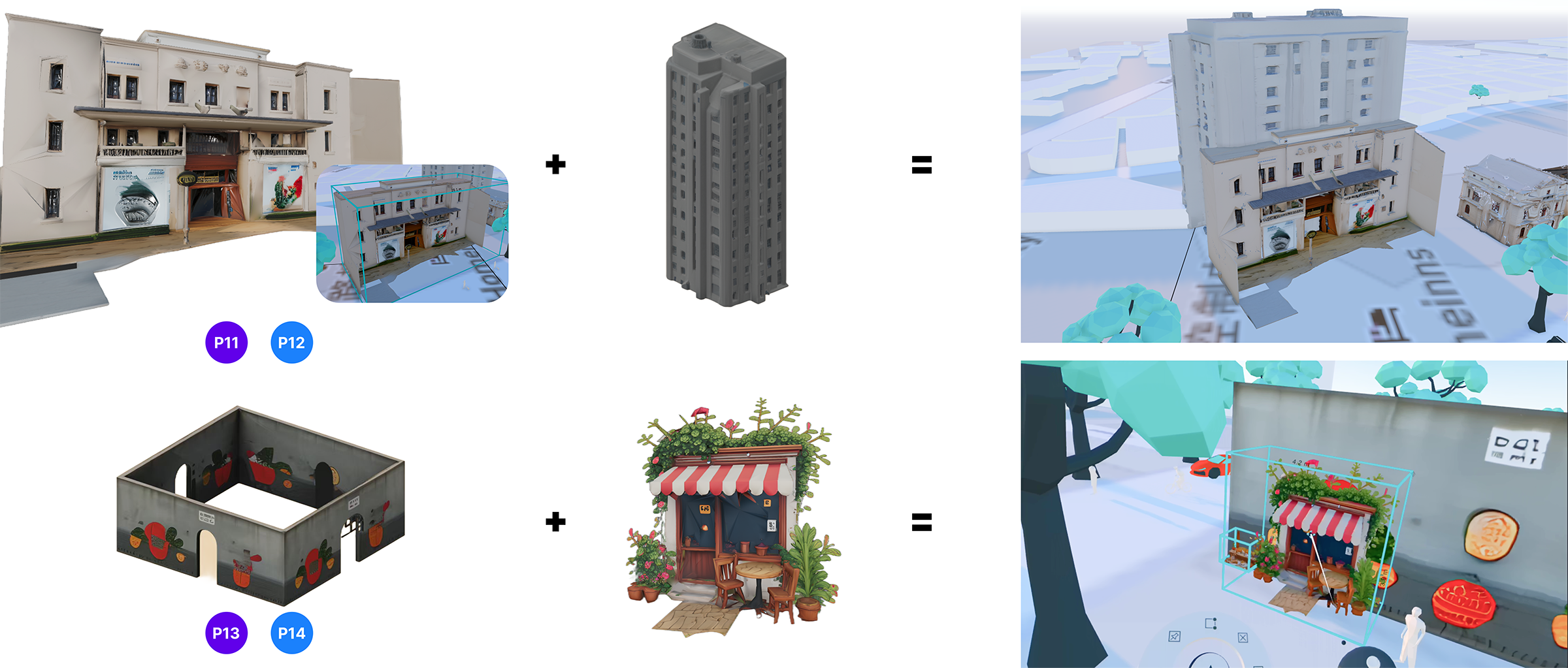}
    \caption{Participants developed a modular, "Lego-like" building strategy. They generated individual architectural components, such as walls, roofs, and windows, then used the system's manipulation tools to assemble them into a complete structure.}
    \Description{Participants developed a modular, "Lego-like" building strategy. They generated individual architectural components, such as walls, roofs, and windows, then used the system's manipulation tools to assemble them into a complete structure.}
    \label{fig:modular}
\end{figure*}

\subsubsection{Creating and Negotiating from Different VR Perspectives}  

Participants strategically switched between perspectives to manage different collaborative tasks. The god's-eye view was used for macro-level planning and division of labor. For instance, from above the virtual Huangpu River, P9 and P10 decided: "I'll fill the west bank with old buildings," and "I'll take the east bank to build the Lujiazui."

Conversely, the first-person perspective was critical for evaluating aesthetic and affective qualities. Standing on a rooftop of a generated building, P5 remarked, "This feels like standing in Pudong... Now I can look down from any building's top," an experience that triggered personal memories. P5 then invited the partner to share this perspective, creating a moment of shared experience. Later, standing at human scale in front of their co-created coffee shop, they affirmed the scene's authenticity: "This is the feeling! This feels exactly like Shanghai." (figure \ref{fig:perspective})

\subsubsection{Developing a Modular Building Method of Generated Models} 

During the experiment, we observed several participants developing a modular building method, assembling structures from AI-generated components in a process similar to using Lego blocks.

P13 and P14 first generated a featureless exterior wall from their memory. They then generated separate architectural details, such as Chinese-style tiled roofs, doors, and windows. Using the system's free placement and reshaping tools, they then combined these components to assemble a complete building. This combination was not always planned; in another case, after moving and testing, P11 and P12 combined two unsatisfied generated objects into one new building, which would be discussed further in section 4.3.4.(figure \ref{fig:modular})

Later, they reflected on this method. One participant stated that it "can restore the geometric relationship of the building in my memory," while another noted that it "gives me more control... than directly entrusting the generation of the entire building to AI."

%fig11
\begin{figure*}[tp]
    \centering
    \includegraphics[width=0.9\linewidth]{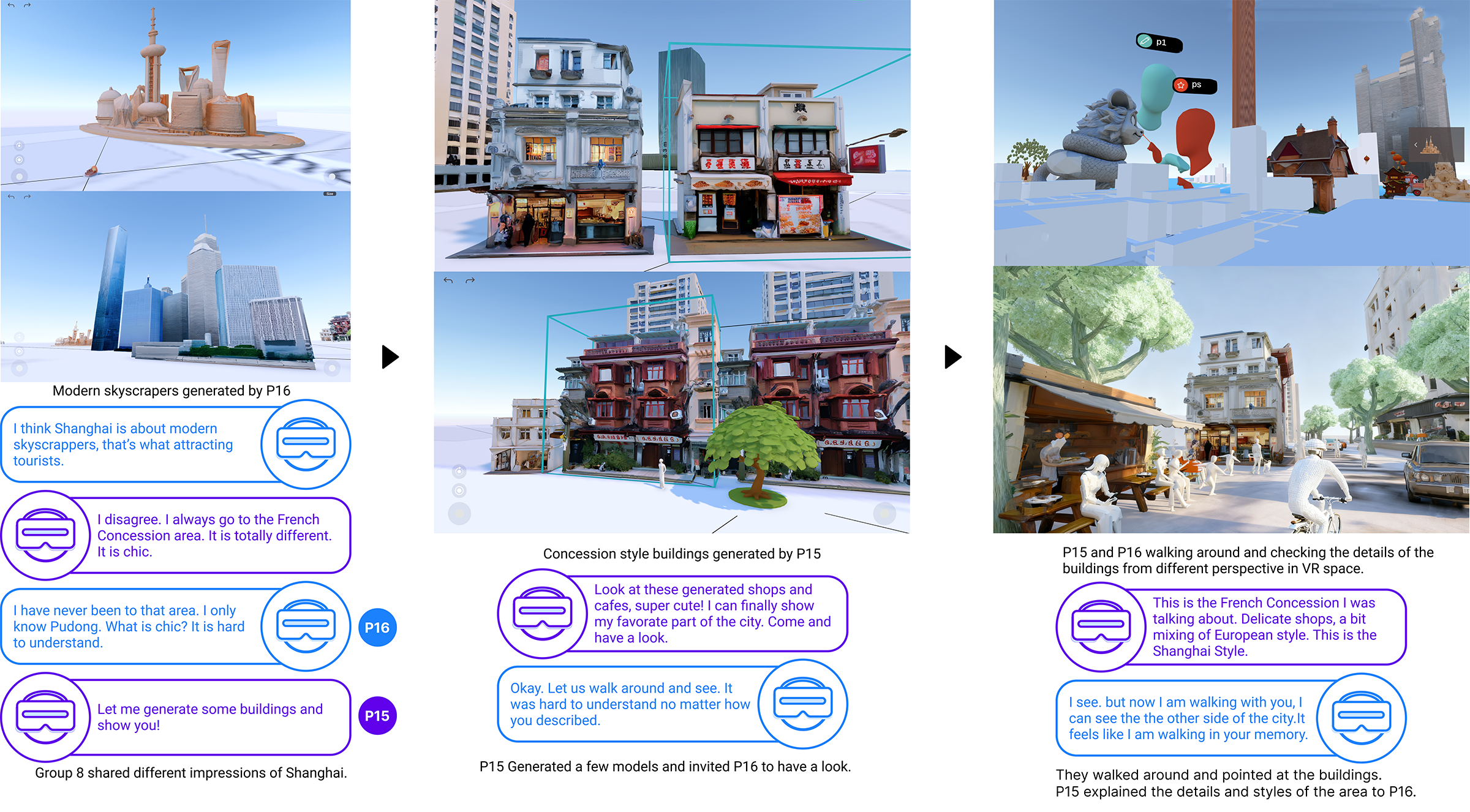}
    \caption{3D models serving as a cognitive bridge. To explain the abstract concept of "French Concession style" to the partner, P15 generated a representative building. By inviting P16 to walk through the shared model, she transformed an abstract idea into a tangible, experiential lesson.}
    \Description{3D models serving as a cognitive bridge. To explain the abstract concept of "French Concession style" to the partner, P15 generated a representative building. By inviting P16 to walk through the shared model, she transformed an abstract idea into a tangible, experiential lesson.}
    \label{fig:gap}
\end{figure*}

%fig12 先留这里看看
\begin{figure*}[tp]
    \centering
    \includegraphics[width=0.9\linewidth]{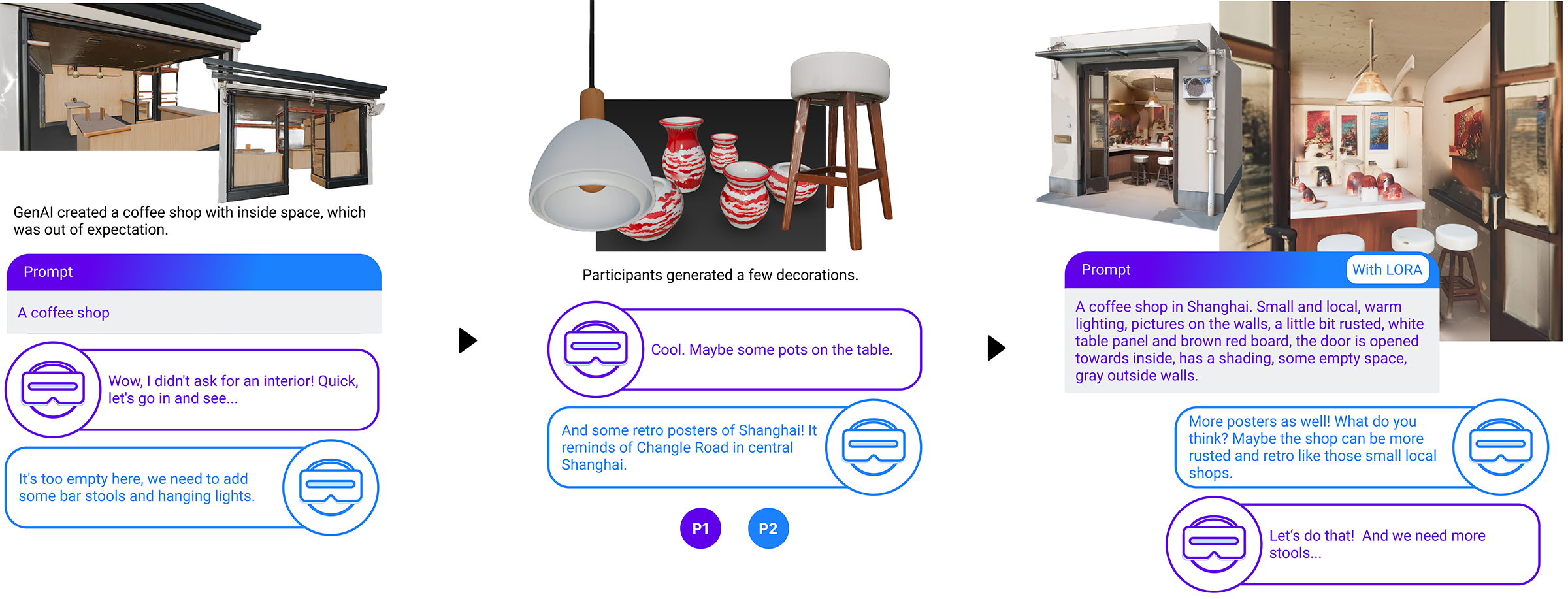}
    \caption{Creative adoption of unexpected AI outputs. When the AI generated an unprompted interior space for a coffee shop, participants did not view it as an error but immediately began collaborating on how to furnish and decorate the new space.}
    \Description{Creative adoption of unexpected AI outputs. When the AI generated an unprompted interior space for a coffee shop, participants did not view it as an error but immediately began collaborating on how to furnish and decorate the new space.}
    \label{fig:interior}
\end{figure*}

\subsection{Challenges of Collaboration: Negotiating Differences Between Participants}   
\subsubsection{Using Generated Models to Negotiate Different Memories}  

When partners had different levels of knowledge about the location, the generated 3D models helped them share information. In one pair, this disparity was pronounced: one participant (P16) primarily associated Shanghai with modern skyscrapers, while the other (P15) possessed deeper knowledge of its historical architectural layers. (figure \ref{fig:gap})

To bridge this gap, P15 generated a 3D model of a building in the style of the former Shanghai French Concession area(1849-1943). The 3D model became a shared object they could both see and discuss, turning an abstract description into something concrete. P15 then invited P16 to walk around and through the model together, using the embodied VR space to point out specific features like the roof shape and windows while explaining their cultural significance. This spatial, experiential demonstration was decisive. P16 later reflected:\begin{quote}
    \textit{"She kept describing the 'French Concession style,' but I couldn't picture it... When she generated that model and I could walk around it, it was like she opened a door and let me walk directly into her memory."}
\end{quote}
For P15, the model provided a way to show rather than explain: "I could finally say, 'See? This is what I was talking about.'"(post-study interview). The 3D model allowed the knowledgeable participant to show, rather than just tell, what they meant. This allowed them to move past a limited understanding and learn from each other.

\subsubsection{Negotiating Different Cultural Preferences} 

Another example involved cultural associations derived from popular media. P11 initially wanted to build scenes inspired by a television drama, while P12 wanted a film scene. Both the drama and the film show the old part of Shanghai, especially when it comes to setting up the urban scenes, but they have contrasting visual art styles. The TV drama evokes the ambition and prosperity of 1990s Shanghai, and the other film describes a 1940s Shanghai, which is associated with darker pre-liberation themes. To resolve this, the pair first generated two rough scenes reflecting each style. After virtually walking through the two contrasting scenes, one participant noted that the scene inspired by the film felt “too narrow and gloomy” in first-person perspective. Following this embodied comparison, they decided together to adopt the aesthetic from the TV drama. They stated that this style better matched their shared goal of creating what they described as a "vibrant and forward-looking" representation of Shanghai.

\subsection{Challenges of Representation: Reconciling with GenAI Outcomes}
\subsubsection{Unexpected Generation} 

The GenAI sometimes generated content beyond user prompts. Within the three-dimensional VR environment, participants would often walk into the new spaces and incorporate these unexpected results into their plans.

For example, when a pair generated a street-corner coffee shop. The AI not only created the building's exterior but also generated an interior space, which was not requested. This surprise led participants to discuss how to decorate the interior. During the collaborative session, P1 exclaimed, "Wow, I didn't ask for an interior! Quick, let's go in and see...". P2 replied: "It's too empty here, we need to add some bar stools and pendant lights." The pair then proceeded to generate the decorations and began furnishing the interior together.

In another case of generating an office building (Pair 3), GenAl added commercial space on the ground floor, even including a small plaza and green beds around it. Participants found this street scene very similar to their memory of Shanghai street and started to expand the street based on this extra space, such as adding local elements like yellow and green shared bicycles. (figure \ref{fig:interior})

%fig13
\begin{figure*}[tp]
    \centering
    \includegraphics[width=0.9\linewidth]{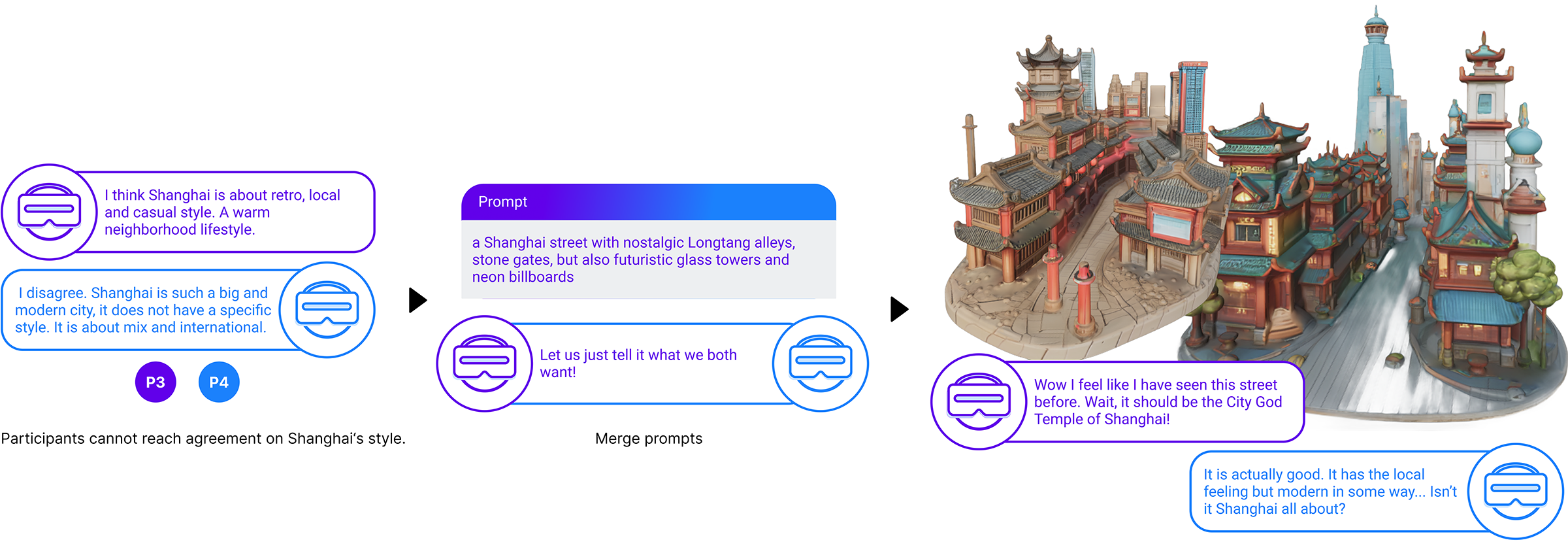}
    \caption{The AI acting as a mediator to resolve conflict. When P17 and P18 disagreed on architectural styles, they co-authored a hybrid prompt. The AI's output, which successfully blended both of their visions, was accepted as a satisfying compromise.}
    \Description{The AI acting as a mediator to resolve conflict. When P17 and P18 disagreed on architectural styles, they co-authored a hybrid prompt. The AI's output, which successfully blended both of their visions, was accepted as a satisfying compromise.}
    \label{fig:hybrid}
\end{figure*}

\subsubsection{Generating for Day and Night}
We observed that the ability to set the virtual time of day and change the lighting prompted participants to generate time-specific elements. For example, in a nighttime setting, P13 generated street lamps. Additionally, P15 and P16 conceived that “a Shanghai street at night should present a vivid and bustling atmosphere filled with colorful lights. Consequently, they generated architectures such as theaters and restaurants adorned with neon and warm lights.

%fig14
\begin{figure*}[tp]
    \centering
    \includegraphics[width=0.9\linewidth]{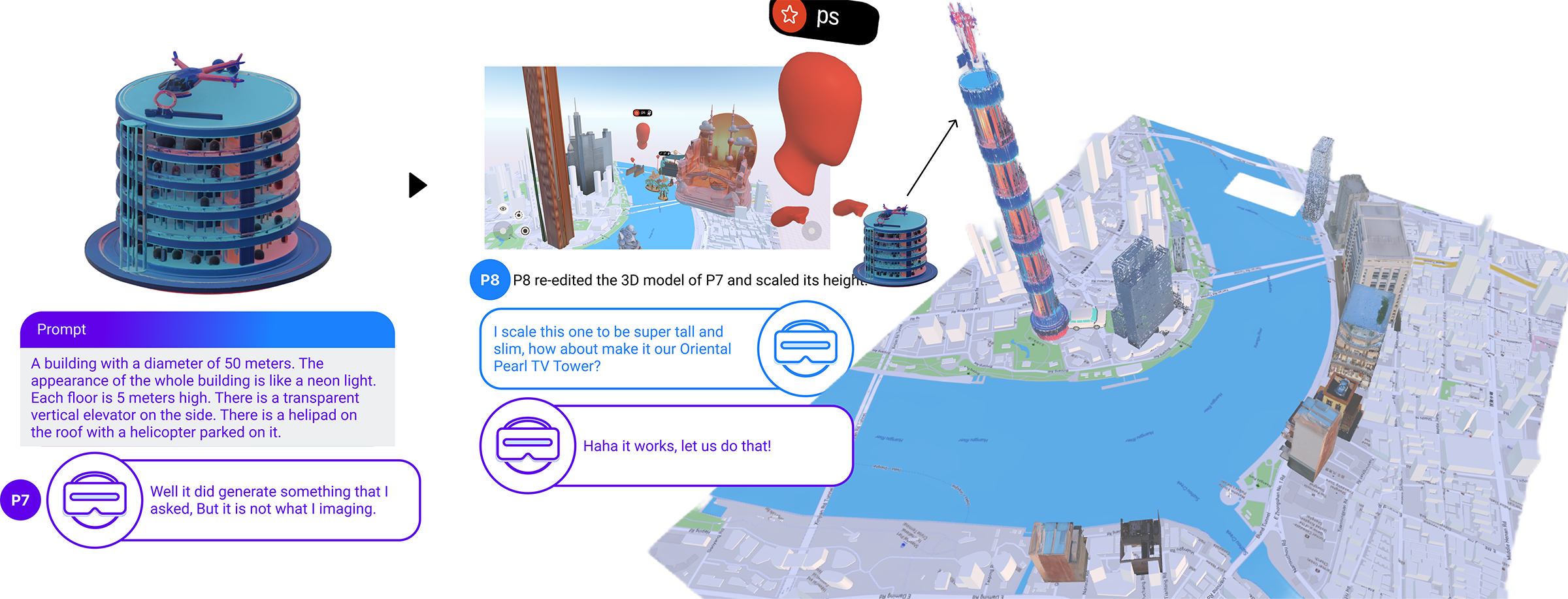}
    \caption{Creative repurposing of a "failed" generation.}
    \Description{Creative repurposing of a "failed" generation.}
    \label{fig:tower}
\end{figure*}

\subsubsection{Hybrid Generation} 

When participants had conflicting style preferences, we observed instances where they used the AI to generate a hybrid compromise. In these cases, participants described the AI's output as "inspiration" that blended both of their ideas. It is neither entirely aligned with one party's requirements nor simply a summation of them, but a result that was described by participants as "inspiration that blended both ideas”.

This was clearly illustrated by P17 and P18, who disagreed on what represented "authentic" Shanghai: one argued for the old, vibrant alleyways ("longtang"), while the other argued for a modern, international metropolis. Unable to agree on their views, they wrote a contrast prompt: "A modern high-rise building with some traditional Chinese elements." The AI's output, a building that seamlessly blended modern and vintage stylistic elements, was accepted by both parties as a successful and satisfying compromise. They placed this building at the entrance to their alleyway and decided to use this hybrid style as a template for further co-creation. When talking about this compromise in the interview, they reflected that the AI's solution “captured and represented Shanghai's embracing spirit”. (figure \ref{fig:hybrid})
%fig15
\begin{figure*}[tp]
    \centering
    \includegraphics[width=0.9\linewidth]{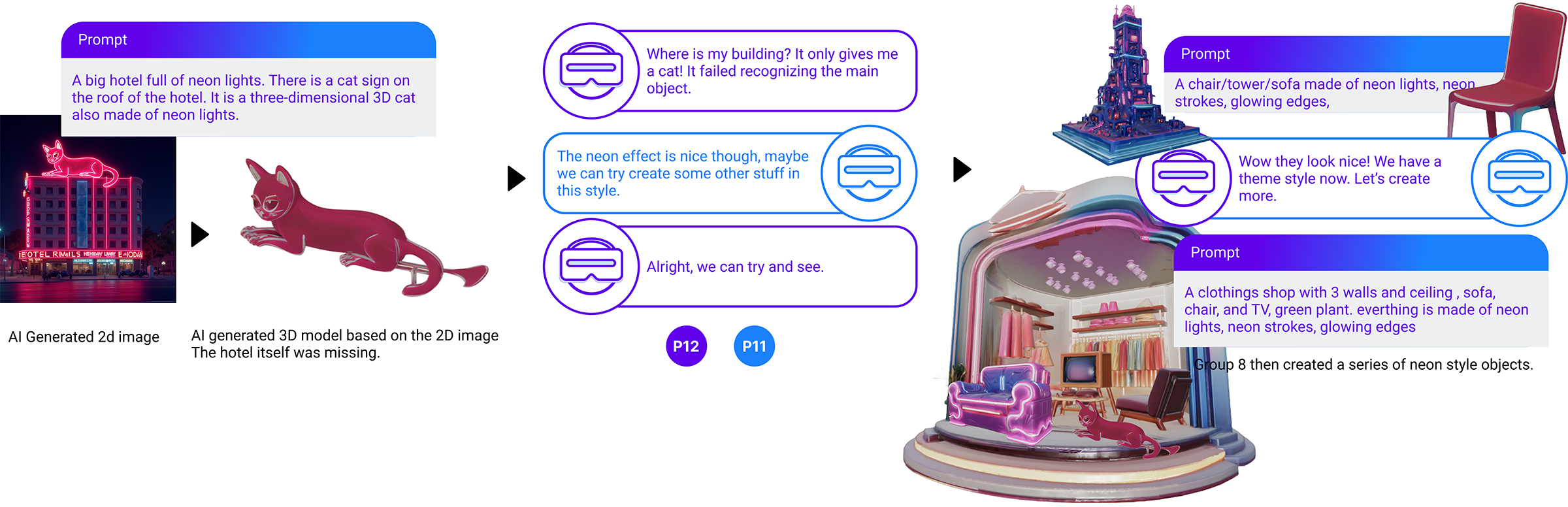}
    \caption{The AI failed to generate a complete building but successfully created a "cat-shaped neon sign" in a style the participants liked. They then adopted "neon style" as a core creative constraint for subsequent generations.}
     \Description{The AI failed to generate a complete building but successfully created a "cat-shaped neon sign" in a style the participants liked. They then adopted "neon style" as a core creative constraint for subsequent generations.}
    \label{fig:neon}
\end{figure*}

%fig16
\begin{figure*}[tp]
    \centering
    \includegraphics[width=0.9\linewidth]{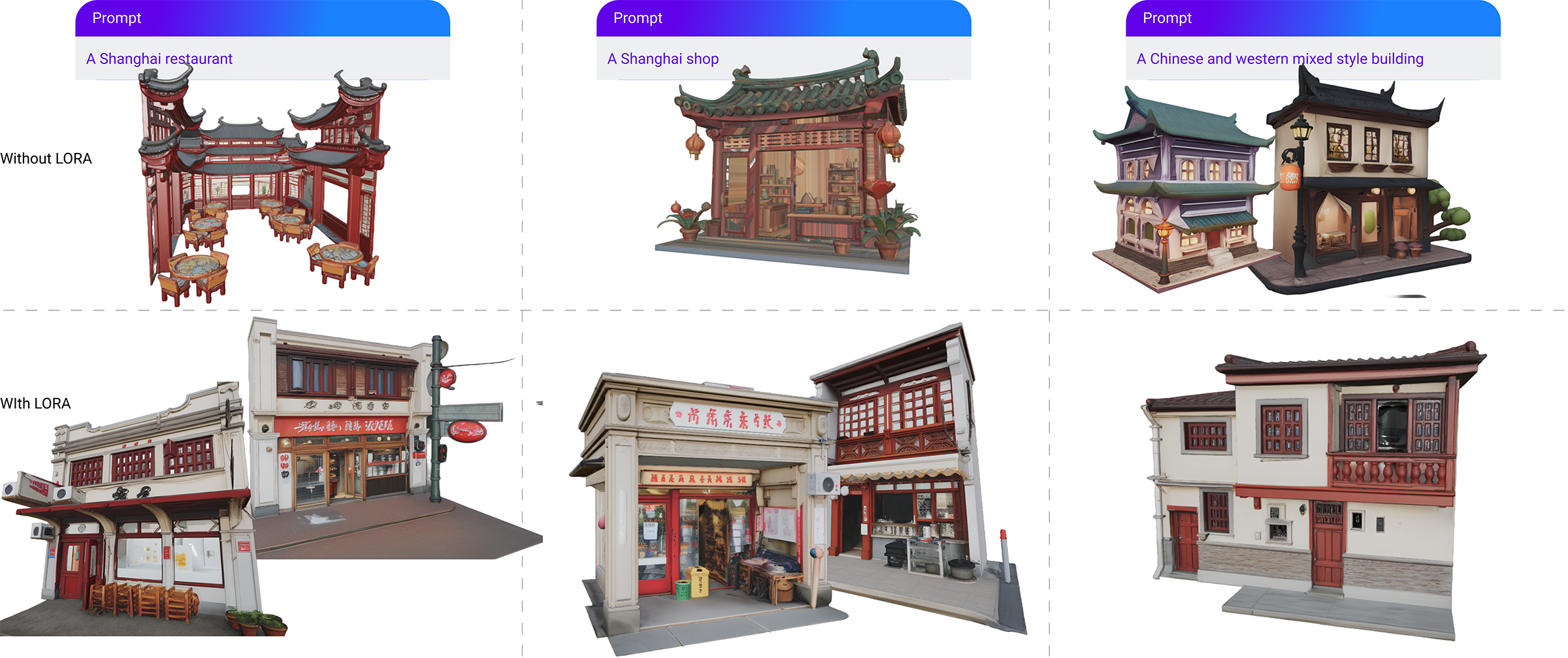}
    \caption{The comparison of using and not using custom LoRA.}
    \Description{The comparison of using and not using custom LoRA.}
    \label{fig:laracompare}
\end{figure*}

\subsubsection{Responding to GenAI’s Shortcomings collectively} 

When GenAI outputs failed to meet expectations, participants did not see it as a dead end. Instead, they collaboratively developed and employed three typical strategies to adapt, repurpose, or work around these limitations, demonstrating their ability to maintain creativity despite AI limitations.

\textit{Repurposing "Failed" Generated Models} One strategy was to find a new use for an object that the AI generated incorrectly. A clear instance of this occurred when a pair attempted to generate a specific architectural feature ("a building with side stairs and a car ramp"). The AI instead produced a structure with illogical columns and a spiral staircase. Rather than discarding it, the pair physically rotated the model, discussed its new potential, and collectively re-narrated it as a "futuristic city installation." They then decided to place it in their street as a central artistic feature. As P10 concluded: "Alright, clearly a car cannot drive up it. But look at it... It doesn't look like a building, more like an installation. ” P9 added: “Let's make it a sculpture then.” In another example, P8 took a model generated by P7, scaled its height, and repurposed it as their virtual city's "Oriental Pearl TV Tower", as seen in figure \ref{fig:tower}.

This strategy was not limited to single objects. As noted earlier (in Section 1.2.3), participants would sometimes regard multiple AI outputs as "components" and combine them to be a new, coherent structure that better matched their intent. This strategy also included unsatisfied outputs. For example, group 6 tried to generate a flower shop, GenAI only returned the front face, without the side and back. They then decided to put this front face onto a gray building, because "the gray building is dull, not like a vibrant Shanghai street that we are picturing.".

\textit{Iterate on successful aspects of failed outputs} A second strategy was to identify one good feature in a failed generation and use it as a starting point for new prompts. Participants would identify a satisfied aspect of a failed output and iterate upon it through subsequent prompts.

For example, P12 requested a "building with a cat-shaped neon sign on the roof." The AI generated only the neon sign, without the building itself. However, the group was satisfied with the visual style of the neon sign. They therefore abandoned the original goal and adopted a new strategy: using "neon style" as the core prompt for future actions, such as generating a "neon style chair." This allowed them to build up a consistent aesthetic atmosphere, despite the AI's initial failure to comprehend the full request. (figure \ref{fig:neon})

\subsubsection{Encountering and Responding to Biased Outputs}

We also observed that without using the custom LoRA, the base GenAI model sometimes generated outputs that participants perceived as stereotypical, as in figure \ref{fig:laracompare}. For instance, when P13 tried to generate "A building with a street lamp that emits warm yellow light... integrating Chinese and Western styles", the model produced a building with elements like "excessively upturned dougong", "lanterns", "overwhelming Chinese red”, and "Chinese horticulture". P13 reacted negatively to this output, commenting: "This is too Western. It looks like a building from the underworld. I don't like this outcome."

%% file: sections/DIS26_05_Discussion.tex
\subsection{3D Spatial Construction Enables Embodied Experience}
Our study reveals an exploration from 2D visualization to 3D narration. Previous studies utilizing GenAI for heritage focused on generating 2D imagery that allows people to visualize a cross-sectional memory. Recent work used GenAI to generate cultural memories and empower the expression of matters of cultural value beyond the physical \cite{Fu2024BeingEroded}. In their work, participants used text-to-image tools to generate 2D visualizations, enabling them to articulate future visions and emotional concerns (e.g., climate anxiety) that are often absent in official archives. We share this rationale for using GenAI to bridge the skill gap \cite{epstein2022happyaccidentssparkcreativity} and subvert the AHD \cite{Smith2006usesofheritage}. However, even when VR technologies enable 360-degree observation of generated imagery \cite{miller2025eliciting,lu2023photovoice}, these experiences remain constrained by 2D or panoramic formats. As a result, participants primarily act as observers, editing a fixed viewpoint rather than actively shaping a space.

Furthermore, compared to recent work \cite{miller2025eliciting}, which utilized GenAI to edit 360-degree panoramic photos, our approach offers different affordances of spatial flexibility. Their approach anchored participants to a fixed viewpoint, providing an effective way of recalling memories triggered by specific visual cues \cite{Douglas2002photoelicitation}. In contrast, our component-based 3D workflow allowed participants to switch perspectives dynamically, from a "god-eye view" to plan the street layout to a "first-person view" from a virtual rooftop. This volumetric freedom not only facilitated the recollection of spatial scenes but also supported the reconstruction of embodied experiences within those spaces, such as navigating narrow alleys. This observation suggests that, for urban heritage, the manipulation of spatial structure is as crucial as the rendering of texture.

%They have real photos, we do not. They have old people, we do not, which will be addressed in 6.5
 Another distinction is that their "photo-elicitation" approach was rooted in real-world captures of specific locations. This visual authenticity provided strong cues for reminiscence, particularly effective for older adults recalling past stories. Although our study employed a 3D environment, the immersive view lacked the photorealistic context of a real location. Hence, in future research, we can examine potential enhancements through augmented reality (AR), as discussed in Section 6.1. Additionally, while their study involved older adult participants, our groups consisted primarily of people in the younger generation who were familiar with technologies, which facilitated smoother interaction with the system interface but excluded valuable perspectives from older generations. The workflow could be extended to explore hybrid methodologies. For example, integrating the visual grounding of real-world panoramas with our 3D construction framework may enable inter-generational storytelling, in which older adults contribute lived experiences of place while younger generations develop dynamic, forward-looking spatial visions. The implications of age-related differences and the potential of observing how narratives are negotiated and transmitted across generations are further discussed in Section 6.5.

%Beyond verifying visual details, our workflow enabled participants to materialize embodied memories that are hard to articulate verbally, specifically the "atmosphere" and personal experience. 
\added{Beyond verifying visual details, it appears that participants utilized the workflow to materialize embodied memories that are hard to articulate verbally, specifically what the participants described as the "atmosphere" and personal experience.} For instance, whereas participants in previous 2D studies might generate an image of a crowded street, our workflow allowed participants to build and walk into their heritage space. They walked the streets not only to observe but also to feel if the width matched their lived experience. This aligns with previous researchers' concept of embodied interaction\cite{Paul2001WheretheActionis}, where meaning arises from physical engagement with the environment. Take the "atmosphere" of the Shikumen lane as an example: by structurally adjusting the density, one pair gave physical shape to the oppressive yet warm atmosphere of a crowded lane (4.1.1). \added{When discussing this spatial adjustment in the post-study interview, P4 explicitly noted the difficulty of verbalizing this feeling: "It was hard to describe the atmosphere, but when we moved them closer, it felt right."} This simple gesture did more than correct a layout; it restored \added{what the participant described} as "intimacy" of the memory. While static images fail to convey this sensation, our participants actively articulated this intangible feeling through spatial manipulation. This confirms that designing CH in immersive space is not just about visual accuracy, but about re-enacting the spatial relationships (density, scale) that define the community's emotional bond to the place. Ultimately, by designing imagined artifacts in the virtual world, people can effectively articulate the heritage of the real world. The authenticity of these user-generated spaces lies not in their architectural precision but in their capacity to express lived experiences, demonstrating that the reality of heritage is defined not only by static documentation but by human connections.

\subsection{Negotiating differences in a spatial environment}
Our findings extend the understanding of how space functions in collaborative heritage storytelling. Recent work explored spatial storytelling in VR \cite{ShinandWoo2023HowSpaceIsTold}, identifying how authors utilize spatial trajectories to guide the audience's narrative experience. Their work, grounded in the concept of spatializing narrative \cite{RYAN2016narratingspace}, conceptualizes the site as a fixed canvas to which narratives are spatially bound for presentation. In contrast, our study uncovers the internal social process between collaborators. In contrast, our study reveals that in a generative co-construction context, space is not a fixed constraint, and spatial actions may serve a social function.

Participants utilized spatial manipulation not only for editing but also as a subtle negotiation tactic. This served to align their shared vision, such as dragging buildings closer to implicitly confirm with each other a desired atmosphere without prompt editing or discussion (4.1.2). It aligns with the distinction between space and place in collaborative systems \cite{Harrison1996Re-place-ingSpace}, where the "place" is constructed not just by geometry but by the social interactions that occur within it. We observed that spatial manipulation was not just about arranging the scene for a viewer but also about delivering their intent to a partner. Unlike text-based collaboration, where disagreement requires explicit verbal negotiation \cite{han2024teams}, our participants used spatial actions as a soft proposal (4.1.2). Rather than conducting verbal arguments, a participant simply pushed a conflicting object to the edge or centered the preferred narrative. In the context of heritage expression, spatial actions can serve as a negotiation tactic; different perspectives and conflicting ideas can settle into a shared arrangement without the social friction of verbal rejection. 

Beyond resolving conflicts, the spatially arranged AI-generated models could function as "boundary objects" \cite{star1989institutional}. We observed this in the French Concession example. When one participant lacked historical knowledge, the other used the generated buildings to create streets as a teaching prop, pointing to them to explain specific cultural nuances and design details. Here, the 3D object bridged a knowledge gap. It turned a potential misunderstanding into a moment of shared learning. The object anchored the conversation, making the abstract description concrete, even as a place where they could walk together and share stories. Ultimately, the value of spatial interaction lies in its ability to smooth conflict and facilitate non-verbal consensus, turning the virtual environment from a mere visualization tool into a social space where people can negotiate their collective memory.

%回应R2关于individual vs collective agency的问题。
\added{The collaborative process also revealed how individual and collective cultural identities interact. When P11 and P12 disagreed on visual style, each first generated their own preferred scene rather than immediately negotiating a compromise. This preserved individual expression before collective decision-making. When P17 and P18 co-authored a hybrid prompt to merge their conflicting visions of ``authentic Shanghai,'' the tools served collective negotiation. The workflow supported both modes as participants retained control over what to generate, where to place it, and whether to accept or revise a partner's contribution.}

%改为这个标题会不会太强烈？
\subsection{The risk of generative AI reshaping heritage memory}
In the context of CH, GenAI introduces significant challenges related to algorithmic bias, lack of tacit knowledge, and semantic misalignment. As noted in AI studies, models trained on broad datasets often default to stereotypical representations, resulting in historical inaccuracies and cultural misrepresentations that fail to capture local nuances \cite{zhang2024partialitymisconception}.

In our study, participants inevitably encountered these algorithmic frictions. One prominent challenge was cultural stereotyping: AI struggled to interpret the specific hybridity of Shanghai style (Haipai), generating "Chinatown" aesthetics or Westernized interpretations that erased local identity, although participants kept adding detailed prompts. Even though we employed a customized LoRA model to enhance local specificity, the system could not cover all cultural perspectives. Beyond the visual style, participants struggled with an atmosphere gap: while the AI could render correct textures (e.g., a brick wall), it was difficult for it to convey prompts of tacit atmosphere directly, such as the "oppressive yet warm" density of a lane. Participants also faced unintended generative artifacts: GenAI often failed to align with user intent, producing visual inconsistencies or non-rational objects, such as buildings defying gravity or chaotic geometries, that physically contradicted the prompt. A more detailed analysis of technical boundaries and implications for future research is provided in Section 6.

%这一段重写了，GenAI在CH语境下会reshape memory。
% 引用Zhou'26。举例neon cat和widening street，重新解读results案例。
Typically, such opacities are viewed as system failures that disrupt the user's flow and threaten the authenticity of heritage representation \cite{Van2011opacity, Sun2024AIhallucination}. \added{ In the context of cultural heritage, this concern carries additional weight. Unlike open-ended design tasks, where unexpected outputs may serve as useful starting points, heritage narratives are grounded in lived experience and personal memory. When GenAI produces outputs that deviate from a participant's intent, there is a risk that the generated content begins to shape the memory rather than represent it. }

\added{Recent work has shown that interacting with GPT during recall tasks can subconsciously alter users' beliefs and perceptions of remembered events \cite{Zhou2026TellMeWhatIMissed}. In our study, we observed instances that echo this concern. For example, when the AI failed to generate a building with a cat-shaped neon sign and instead produced only the sign, the pair abandoned their original goal and adopted ``neon style'' as a new creative direction (Section 4.3.4). While this pivot produced an aesthetically coherent scene, it also meant that the resulting narrative was driven by the AI's output rather than by the participants' recollections. Similarly, when a pair widened their virtual street and then decided to build a plaza, the design intent shifted from reconstructing a remembered place to responding to what the tool made available. These moments illustrate how the generative process can redirect the narrative away from memory and toward what the AI happens to produce.}

\added{At the same time, these observations point to a broader characteristic of memory itself. Memory is not a fixed record; each act of recall is an act of reconstruction \cite{Jin2022fluidheritage, Brady2017memoryconstructive}. When participants responded to unexpected AI outputs, they were doing what people do with memory in everyday life: adjusting, filling gaps, and revising the narrative based on new information. As a recent work illustrated through a physical-digital archive that uses GPT-based interpretations of a city's history, our remembrances of the past are always shaped by the technologies and biases of the present, and personal memories are subject to reinterpretation and change over time \cite{LC2025ArchiveFuture}. The co-creation process in our study made this mutability visible. For instance, P16 initially could not picture the ``French Concession style,'' but after walking through a generated model together with P15 (Section 4.2.1), P16's understanding of Shanghai's architectural history was reshaped through the interaction. This does not diminish the concern about AI reshaping memory. It means that heritage systems should be designed with this risk in mind, providing participants with tools to reflect on how their narratives are formed rather than assuming that the output faithfully mirrors what was remembered. By making the reconstruction process explicit and shared, the system could allow participants to see where their memories align, where they diverge, and where they are influenced by external input, whether from a partner or from the AI.}

%增加段落
\subsection{Cultural specificity and the limits of universal models}
%homogenization, 2 ref papers, 1AC, 2AC
The friction participants encountered with the base model are not isolated technical failures. They reflect a structural problem in how generative models represent culture. Researchers have argued that large language models have become drivers of cultural homogenization, operating at a scale that exceeds previous technologies, and that a feedback loop forms when AI-generated content becomes training material for future systems, progressively narrowing cultural representation \cite{Daryani2026homogenizingengine, Bender2021stochasticparrots}. A model trained on globally aggregated data will systematically underrepresent local cultural forms, shaping what participants can express and what remains out of reach \cite{zhang2024partialitymisconception}. While their analysis addresses the homogenizing tendency at a systemic level, other works provide empirical evidence of how this plays out in practice: in a controlled cross-cultural experiment, AI writing suggestions led Indian participants to adopt Western writing styles, altering not only the content but also the manner of expression \cite{Agarwal2025homogenize}. Together, these studies show that Western-centric AI models do not merely produce inaccurate outputs for non-Western users; they actively pull users' expressions toward dominant cultural norms. Our study extends this concern to the domain of visual heritage: when participants used the base model to generate Shanghai architecture, the outputs defaulted to stereotypical representations of Chinese aesthetics rather than capturing the specific hybridity of Haipai style. The model's cultural assumptions and identity~\cite{huang_not_2026} might affect what participants could express before they even began to negotiate their memories.

%contextual local, LoRA(respond to 1AC,R3)
\added{The performance difference between the base model and the custom LoRA model in our study supports a growing argument in AI research for locally bounded models, trained on the cultural and aesthetic knowledge of a specific community, rather than relying on a universal standard model \cite{Qadri2025nonwesternartworlds}.  When the LoRA model understood local architectural language, participants could direct their cognitive effort toward memory and narrative rather than correcting stylistic errors. For example, P18 used a simple prompt, and the LoRA model generated a two-floor building with a retro Shanghai style that met their expectations, allowing the pair to build upon this output with more specific details (Section 4.1.1) and to engage more directly with heritage narratives.}

%% file: sections/DIS26_06_Limitation.tex
\subsection{Limitations of Ecological Validity in Virtual Heritage}
Our study was conducted in a controlled VR environment, which, while necessary for the experiment, is not ecologically fully valid in representing how heritage discussions naturally occur. In the real world, collaborative remembering is closely tied to the surrounding environment. The physical scale of a building, the ambient sounds of a neighborhood, or the texture of a wall are all powerful sensory cues that trigger memories and shape conversation. By conducting the study away from the actual heritage site, participants could not draw on these immediate sensory details as a shared reference for their generative prompts. For example, the way participants pointed to and discussed a digital model on a virtual table is fundamentally different from how they might gesture towards a real building while standing on the street. This aligned with findings that in heritage-related VR, satisfaction is mainly driven by "activity-related authenticity", the satisfaction of heritage-related experiences depends more on the authenticity of "situations/activities" \cite{nam2023authenticity}. These findings further confirm the ecological validity limitations of VR relative to the real site. 

Future work could address this by using Augmented Reality (AR). A follow-up study could involve participants walking through the actual site, with AR used to overlay and manipulate AI-generated models directly in the real-world environment. This would allow us to observe how real-world landmarks and the physical presence of other people influence the prompts participants create together.

\subsection{The Representational Fidelity of Generative Models}
%lacking finer details, partial and incomplete,
Our study relied on the current capabilities of GenAI, and the resulting 3D models, while stylistically appropriate, could not fully replicate the realistic details of real-world architecture. We used a fine-tuned LoRA model to ensure the generated buildings aligned with the Haipai architectural style, capturing its core structural and formal elements to some extent. However, the finer details, such as the texture of brickwork, the reflection on a window pane, or the subtle rust on the facade, were often rendered with a degree of abstraction or blurriness. Additionally, the generative 3D pipeline introduced issues of geometric fidelity \cite{Zhao2025RevivingMuralArt,poole2022dreamfusiontextto3dusing2d}.  Current image-to-3D models heavily rely on the visible information in the input images and have difficulty in inferring and completing the unseen parts on prior knowledge \cite{PANG2022102859}, such as architectural style and symmetry, like the back facades and roof structures of “lilong” houses. As a result, the generated 3D models are often partial and incomplete; current methods cannot fully reproduce the authenticity or perceived 'aura' of original works \cite{Zhao2025RevivingMuralArt,Benjamin1968WorkofArt}. Our participants, in essence, negotiated and co-created with a simplified, somewhat idealized version of their memories rather than with a photorealistic replica. 
%could impact storytelling
This gap between the generated representation and physical reality could have influenced the collaborative process. For instance, the lack of fine-grained, realistic detail could have made it easier for participants to merge their differing perspectives, as there were fewer concrete, high-fidelity features to contest. Conversely, it might have also limited the depth of memory recall, as the specific sensory details that often trigger memories were absent \cite{buchanan2007Retrievalofemotionalmemories}.

\added{Future research should consider investigating how model fidelity affects collaborative memory construction. A comparative study, for example, could contrast the collaborative patterns observed using generative models (like ours) with those observed using high-fidelity, photogrammetry-based digital twins of the same heritage sites. This would help reveal how the level of realism in a shared artifact either constrains or enables the negotiation of collective memory.}

\subsection{GenAI Model Lacking Specific Cultural understanding}

When generating historical buildings, the general GenAI model has a limited understanding of local architectural terms in Shanghai. During the study, the model failed to generate key elements in the prompt, such as specific architectural styles like "arcade-style buildings" and "Shanghai-style", or local materials like "granitic plaster" and "red brick jointing". The resulting models contained inaccuracies or stylistic errors and did not match what the participants described, and they couldn't precisely follow the real intention of the participants. For instance, the base model couldn't accurately understand the participant's prompt of "Shikumen-style" residential buildings. To resolve this issue, we have prepared a LoRA model for participants to choose to use. However, our LoRA only targeted the Haipai style of the specific area, while it did not represent all architectural styles in Shanghai and cover all the ideas of the participants. Future work involves developing a broader library of localized models to ensure the accuracy of culturally specific content.

\added{Future work should explore participatory processes for developing locally tuned models, where communities themselves have agency over dataset curation and the aesthetic decisions encoded in the model. Extending the workflow to other cities with layered cultural identities, particularly contexts combining distinct cultural traditions in everyday architecture and urban life, would also help identify which findings are specific to the Shanghai case and which point toward transferable design principles for culturally situated VR heritage systems.}

\subsection{Scalability to larger group configurations}

The current study utilized a two-person workflow, which enables a direct dialogue and negotiation. While effective for observing fundamental collaborative mechanisms, these patterns may not scale linearly to larger groups. In triads or larger teams, the dynamics of interaction may be more complicated, and the increasing group size introduces distinct behavioral modes and conflict resolution strategies\cite{Zajkowski2024groupcooperation}. Situations like coalition, majority voting, or silence-driven consensus may arise; the co-created cultural scenes might differ. The role of GenAI might also evolve from a "mediator" to a tool leveraged by subgroups to enforce a specific design decision agenda.  Future work should explicitly examine these polyadic interactions to understand how social hierarchies and leadership roles influence the collective prompting process.

\subsection{Homogeneity of participant perspectives}

Our participants consisted primarily of digitally literate young adults. While this reduced technical friction during the VR tasks, it inevitably constrained the breadth of cultural narratives. Heritage meaning is inherently multivocal; a long-term elderly resident and a younger resident will construct vastly different interpretations of the same site based on their distinct lived experiences. The absence of older generations is particularly critical, as their narratives often differ from those of the digital native generation. 

Since intergenerational transmission is a core mechanism of cultural continuity \cite{Wang2024intergenrationaliICH}, future research can move beyond peer-to-peer collaboration to include intergenerational pairs, such as mid-age and young adults, grandparents and grandchildren. This may reveal how cultural story is not just agreed upon, but retrieved, contested, and maybe passed down across temporal divides. Furthermore, addressing this demographic gap requires reconsidering the accessibility of the workflow; future systems might explore accessible interfaces, or a mixed way that a younger user operates the VR while an older user guides, to leverage the strengths of both generations.

%% file: sections/DIS26_07_DI.tex
\subsection{Enabling multi-scale narratives of external and internal spaces}
In our study, when GenAI generated interior spaces, participants shifted their attention from the building's facade to the internal living spaces, entering to observe and even designing them to convey richer personal experiences. However, current generative tools for virtual reality heritage typically default to solid objects. For themes as closely connected to living experience as cultural heritage, the design system should support multi-scale interaction. It should not merely generate a cube, but also support spatial exploration. It can provide users with specific spatial prompts to assist in generating inner spaces, and it can also utilize the freedom of VR space, which allows users to seamlessly switch between the "macro" scale of the street and the "private" scale of a room. This would support the complementary narratives between public heritage sites and private memories, capturing the intangible experiences that constitute community memory.

\subsection{Supporting Component-Based GenAI Workflow in VR}
We observed that participants spontaneously adopted a Lego-like construction strategy, piecing together elements (walls, doors) rather than generating a whole scene at once (4.1.4). This aligns with the reconstructive nature of memory, where details are often recalled in fragments \cite{Brady2017memoryconstructive}. Design tools can explicitly support this workflow by offering component-based generation modes. Design interfaces should allow users to decompose complex prompts into separate assets and provide precise spatial assembly tools to reduce the technical friction of assembly, such as snapping, scaling, and grouping within VR. This grants users greater control in reconstructing their memories, rather than accepting the whole scene.

\subsection{Visualizing embodied negotiation signals}
Our study highlights that users may resolve cultural disagreements through silent spatial manipulation rather than verbal debate(Discussion 5.1). To support this, we can make interactions more obvious in collaborative VR systems. For instance, the system could introduce a negotiation preview mode in which User A moves User B’s object, while both the original position and the proposed change remain visible as semi-transparent elements until the change is mutually accepted. By treating manipulation as a visible communicative act, the system may help reduce the friction of verbal argumentation and support a smoother co-creation process, essential for handling sensitive, contested heritage narratives.

\subsection{Bridging Design Gaps through a culturally diverse Model Library}

We found that using a culturally specific LoRA (Shanghai style) helped smooth the co-creation process, allowing participants to focus on narratives rather than correcting stylistic details. Users often lack the design vocabulary to trigger these styles via text, but the GenAI base models often lack the understanding of specific local contexts. Future systems can introduce style-aware interfaces with a library of fine-tuned models that represent various local aesthetic and cultural elements. Instead of solely relying on text prompts, the system bridges the gap between expert design knowledge and layperson expression. By enabling users to select visual tags that the system translates into style-specific model weights, the approach reduces cognitive burden and ensures culturally appropriate outputs, allowing the community to focus on shared memories. \added{A related study using LoRA fine-tuning for Chinese intangible cultural heritage found that while stylistic accuracy improved, the cultural meaning of generated outputs remained insufficient without ongoing community involvement in the design process\cite{Yuan2025huayao}. In our study, the LoRA model was assembled by the research team, encoding choices about which buildings and styles to include. Future systems should explore participatory approaches to curation, so that the model reflects the community's own understanding of their heritage.}

%censorship, R2
\subsection{Considering guardrails and transparency in heritage AI systems}
\added{A related concern is that in some regions, the definition of cultural heritage may be shaped by political discourse\cite{Smith2006usesofheritage}. Training data reflects curatorial decisions about what counts as culturally significant, and models may be subject to content restrictions that alter heritage representations\cite{zhang2024partialitymisconception}. In such cases, participants may find that certain memories cannot be expressed because the model has been trained or filtered to exclude specific narratives. For heritage systems that aim to surface community-held stories outside Authorized Heritage Discourse\cite{tsenova2020unauthorised, Claisse2020craftingcritical}, future work should consider safeguards such as allowing communities to audit the training data used in local models and making the model's cultural assumptions transparent.}

%% file: sections/DIS26_08_Conclusion.tex
 This work suggests that CH is not a fixed physical form, but a continuous narrative process that can be actively co-shaped by people; the negotiation itself becomes a form of heritage expression. This mode of AI-driven design transforms heritage storytelling into a spatial discourse. Whether performed through spatial manipulation of virtual artifacts or the responses to unexpected GenAI outputs, these actions serve as instrumental mechanisms for collaborative design. Ultimately, this study demonstrates that design is a process that unlocks people's ability to elicit and surface their relationships with the CH of places in their lives, actively shaping their collective heritage together.

%% file: sections/Appendix.tex
\subsection{Behavior Coding Table}

\begin{figure*} [!b]
    \centering
    \includegraphics[width=1\linewidth]{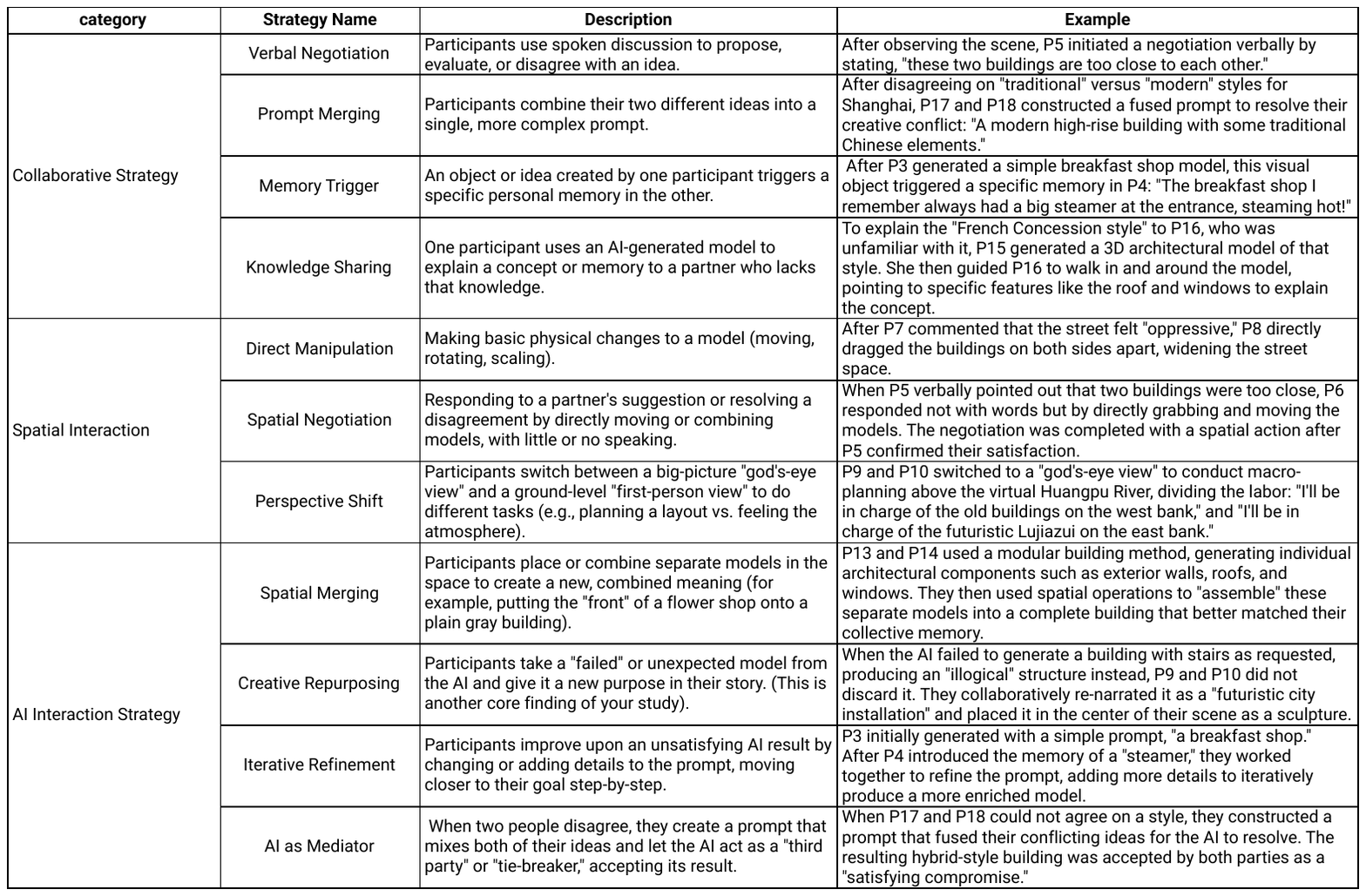}
    \caption{Behavior Coding Table}
    \label{fig:coding}
    \Description{Behavior Coding}
\end{figure*}

The behavior coding based on participants' reactions is shown in figure \ref{fig:coding}.

\subsection{Generated 3D Contents Table} \label{3D Contents Table}

The following table provides a partial log of the generative process from our study, documenting a selection of the prompts authored by participants and the corresponding 3D models produced by the GenAI system. This table presents a representative sample of 79 items.

The log is organized into four categories based on a post-study analysis of the models' styles: "Shanghai Style," "Modern Style," "Hybrid Style," and "Misc. Objects" for items that do not fit the other classifications.

\begin{figure*} [tp]
    \centering
    \includegraphics[width=0.65\linewidth]{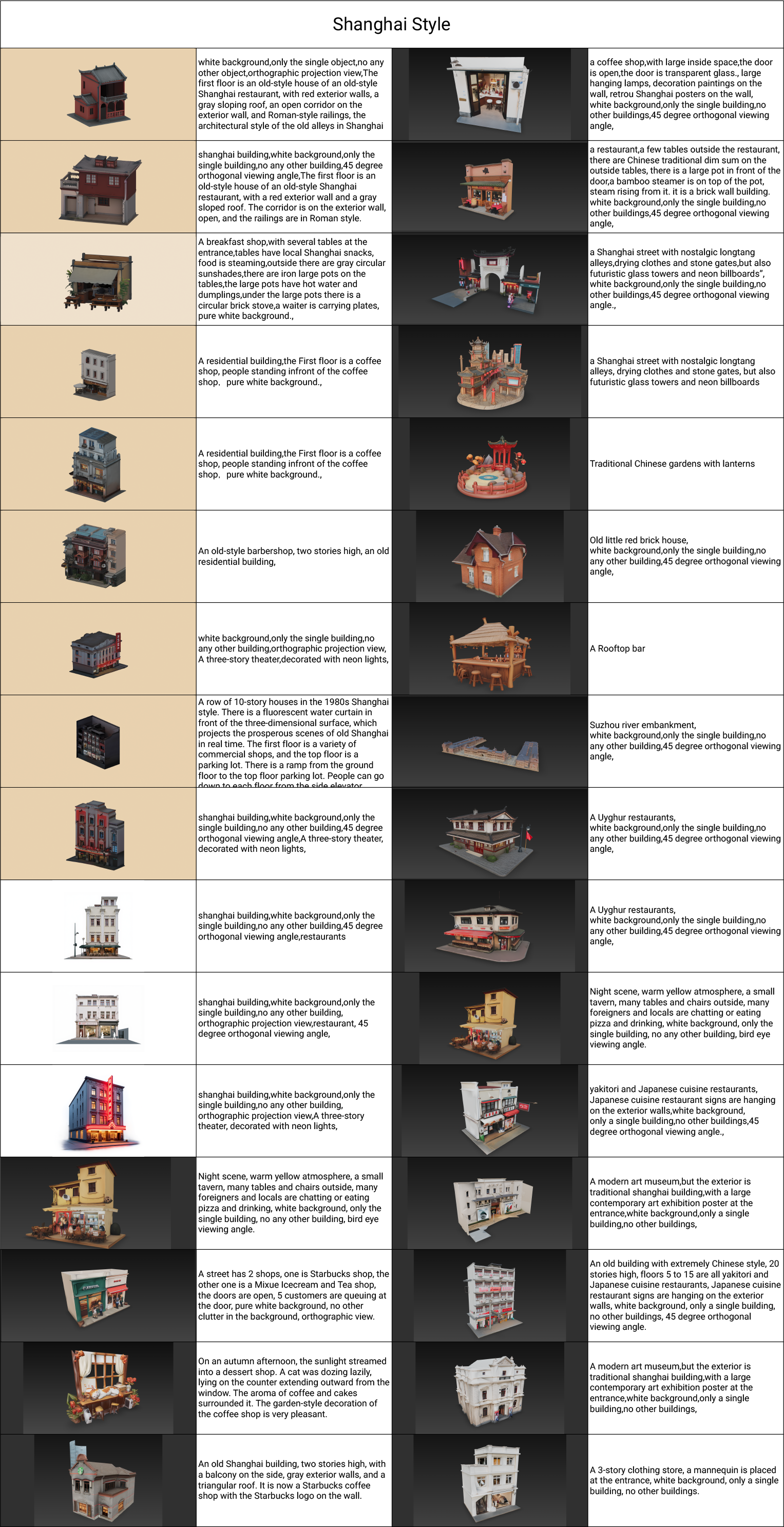}
    %\caption{Your caption here.}
    \label{fig:Content Table 1}
    \Description{Content table 1}
\end{figure*}

\begin{figure*} [tp]
    \centering
    \includegraphics[width=0.65\linewidth]{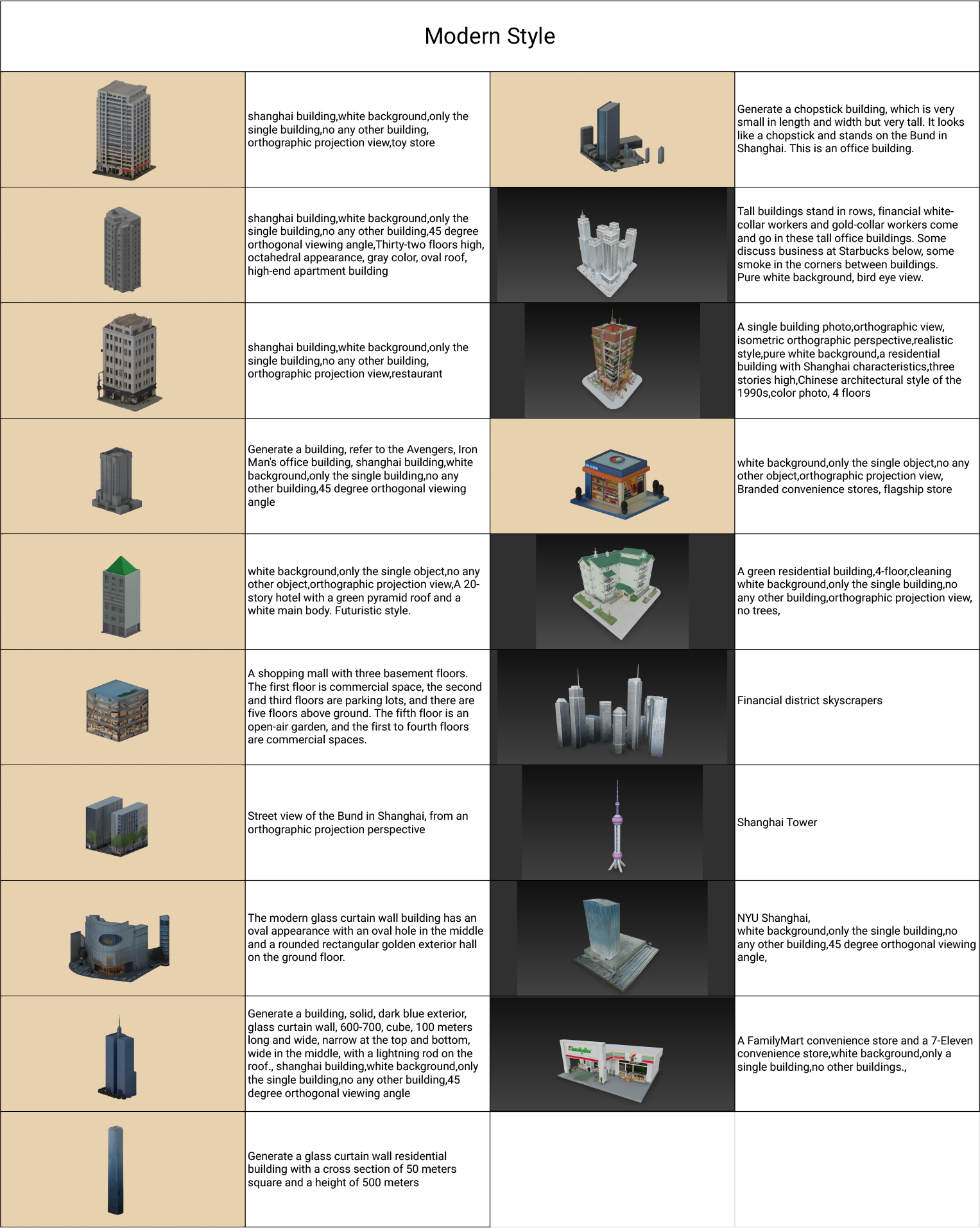}
    %\caption{Your caption here.}
    \Description{Content table 2}
    \label{fig:Content Table 2}
\end{figure*}

\begin{figure*} [tp]
    \centering
    \includegraphics[width=0.65\linewidth]{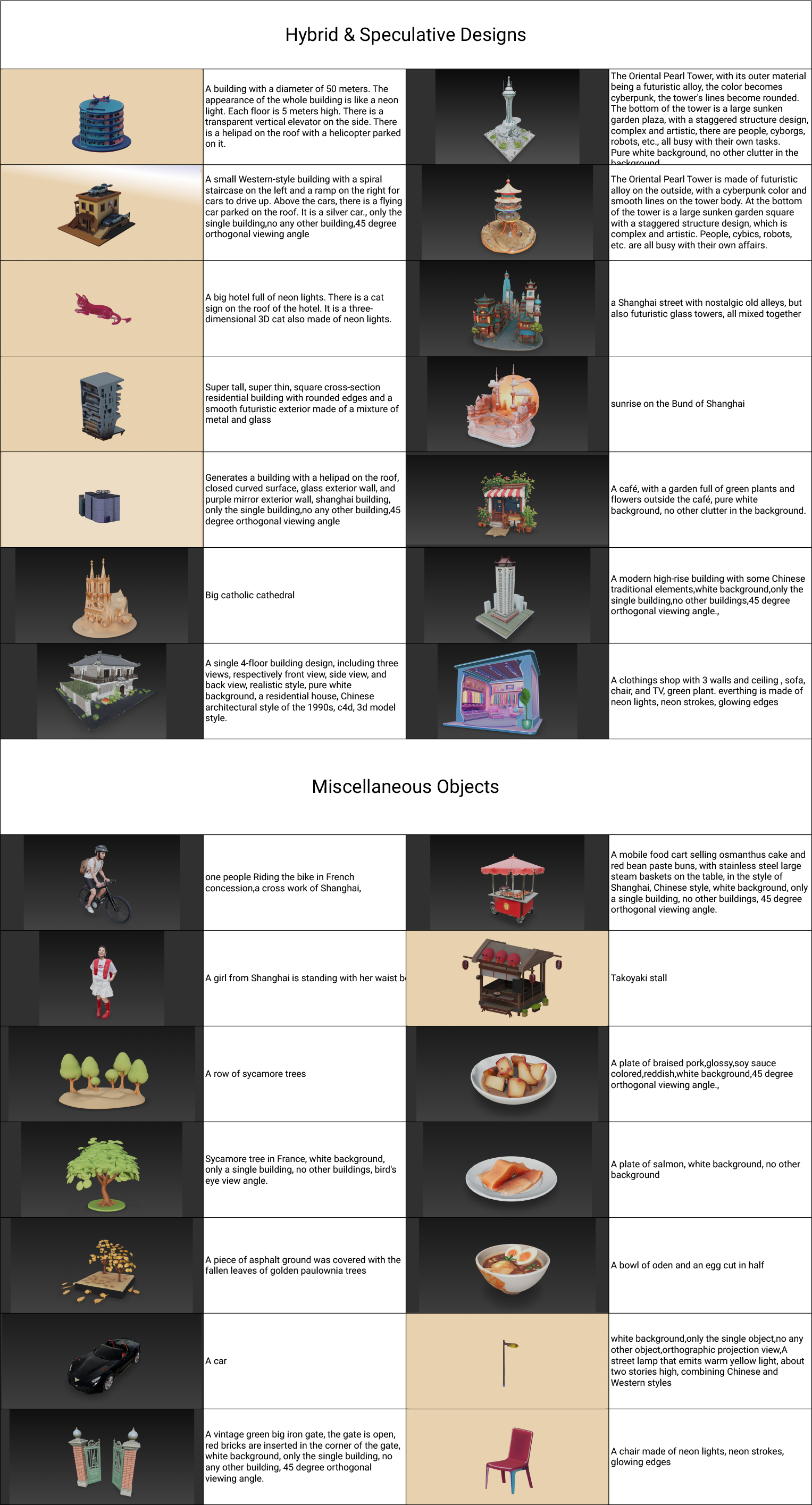}
    %\caption{Your caption here.}
    \label{fig:Content Table 3}
    \Description{Content table 3}
\end{figure*}

% \begin{figure} [htp]

%     \centering
%     \includegraphics[width=0.9\linewidth]{figures/Pre-Study Demographic Questionnaire.pdf}
%     %\caption{Your caption here.}
%     \label{fig:pre study Table 3}
%     \Description{pre study table3}
% \end{figure}

% \begin{figure} [htp]

%     \centering
%     \includegraphics[width=0.9\linewidth]{figures/Post study protocol.pdf}
%     %\caption{Your caption here.}
%     \label{fig:Post study protocol}
%     \Description{Post study protocol}
% \end{figure}